\documentclass[prl,letter,twocolumn,nopacs,superscriptaddress,notitlepage]{revtex4-2}
\usepackage[utf8]{inputenc}
\usepackage[american,]{babel}
\usepackage[T1]{fontenc}
\usepackage[pdftex]{graphicx}  
\usepackage{graphicx, xcolor}
\usepackage{dcolumn}
\usepackage{bm}
\usepackage{amsmath,amsthm,amssymb}
\usepackage{hyperref}
\usepackage[T1,T2A]{fontenc}
\usepackage{xcolor}
\hypersetup{colorlinks,bookmarksopen,bookmarksnumbered,
    citecolor=blue,
    linkcolor=blue,
    pdfstartview=false,
    urlcolor=blue}
\usepackage{graphicx}
\usepackage{braket}
\usepackage{wrapfig}
\usepackage{soul}
\usepackage{mathtools}
\usepackage{float}
\usepackage{tabularx}

\renewcommand{\vec}{\mathbf}

\graphicspath{{figs/}}

\begin{document}
\newcommand{\titleinfo}{Purification Timescales in Monitored Fermions}
\title{\titleinfo}

\author{Hugo L\'oio}
\affiliation{Laboratoire de Physique Th\'eorique et Mod\'elisation, CNRS UMR 8089,
CY Cergy Paris Universit\'e, 95302 Cergy-Pontoise Cedex, France}
\affiliation{CeFEMA, Instituto Superior Técnico, Universidade de Lisboa,
Av. Rovisco Pais, 1049-001 Lisboa, Portugal}
\author{Andrea De Luca}
\affiliation{Laboratoire de Physique Th\'eorique et Mod\'elisation, CNRS UMR 8089,
CY Cergy Paris Universit\'e, 95302 Cergy-Pontoise Cedex, France}
\author{Jacopo De Nardis}
\affiliation{Laboratoire de Physique Th\'eorique et Mod\'elisation, CNRS UMR 8089,
CY Cergy Paris Universit\'e, 95302 Cergy-Pontoise Cedex, France}
\author{Xhek Turkeshi}
\affiliation{JEIP, USR 3573 CNRS, Coll\`ege de France, PSL Research University,
11 Place Marcelin Berthelot, 75321 Paris Cedex 05, France}

\begin{abstract}
We investigate the crucial role played by a global symmetry in the purification timescales and the phase transitions of monitored free fermionic systems separating a mixed and a pure phase. Concretely, we study Majorana and Dirac circuits with $\mathbb{Z}_2$ and U(1) symmetries, respectively. 
In the first case, we demonstrate the mixed phase of $L$ sites has a purification timescale that scales as $\tau_P\sim L \ln L $. At $1\ll t\ll \tau_P$ the system attains a finite residual entropy, that we use to unveil the critical properties of the purification transition. 
In contrast, free fermions with U(1) manifest a sublinear purification timescale at any measurement rate and an apparent Berezinskii-Kosterlitz-Thouless  criticality. We find the mixed phase is characterized by $\tau_P\sim L^{\alpha(p)}$, with a continuously varying exponent $\alpha(p)<1$.
\end{abstract}

\maketitle
\textit{Introduction.---}
Preparing pure states in many-body systems is fundamental for quantum simulation~\cite{georgescu2014quantumsimulation,ferris2022quantumsimulationon,fraxanet2022comingdecasesof}, metrology~\cite{pezze2018quantummtrologywith,desaules2022extensivemultipartiteentanglement,dooley2021robustquantumsensing,dooley2018robustquantumsensing,dolay2023entanglementenhancedmetrology}, and computation~\cite{preskill2018quantumcomputingin}. 
As a process, purification is regulated by the principles of thermodynamics. 
Left alone, quantum systems thermalize with the surroundings and reach a mixed state with extensive thermodynamic entropy~\cite{deutsch1991quantumstatisticalmechanics,srednicki1994chaosandquantum,Srednicki1999theapproachto,Rigol2008thermalizationandits,polkovnikov2011nonequilibriumdynamicsof,foini2019eigenstatethermalizationhypothesis,pappalardi2022eigenstatethermalizationhypothesis}. 
The second law implies that purification requires the system to non-unitary interact with an environment, while the third principle states that the time needed to prepare a zero entropy (pure) state diverges with the system size. 
This purification timescale is in principle dictated by the microcopic properties of the framework. Yet, it manifests a universal behavior, with the functional dependence on system size fixed by the system's dynamical phases.
For instance, generic monitored many-body systems distinguish between mixed and pure phases, with purification timescale exponential and logarithmic in system sizes, respectively~\cite{fisher2022randomquantumcircuits,gullans2020dynamicalpurificationphase,gullans2020scalableprobesof,lunt2021measurementinducedcriticality,sierant2022dissipativefloquet}. Additionally, the purification timescale reveals a power law divergence at the so-called measurement-induced phase transitions (MIPT) governed by the underlying critical field theory~\cite{zabalo2020criticalpropertiesof,sierant2022measurementinducedphase}.
Overall, the  properties of generic systems reflects in the phenomenology of entanglement propagation and transitions~\cite{lunt2022quantumsimulationusing,potter2022entanglementdynamicsin,rossini2021coherentanddissipative,chan2019unitaryprojective,li2019measurementdrivenentanglement,skinner2019measurementinducedphase,czischek2021simulating,han2022entanglementstructure,minoguchi2022continuousgaussianmeasurements,altland2022dynamicsofmeasured,fuji2020measurementinducedquantum,jian2021yangleeedge,jin2022KPZ,willsher2022measurementinducedphase,pizzi2022bridgingthegap,lyons2022auniversalcrossover,zhang2020nonuniversalentanglementlevel,zhang2021emergentreplica,zhang2022universalentanglementtransitions,zhou2021nonunitaryentanglementdynamics,bentsen2021measurementinducedpurification,yang2022entanglementphasetransitions,rossini2020measurementinduceddynamics,medina2021entanglementtransitionsfrom,lunt2020measurementinducedentanglement,kelly2022coherencerequirementsfor,szyniszewski2019entanglementtransitionfrom,szyniszewski2020universalityofentanglement,tang2020measurementinducedphase,iadecola2022dynamicalentanglementtransition,odea2022entanglementandabsorbing,ravindranath2022entanglementsteeringin,piroli2022trivialityofquantum,sierant2022controllingentanglementat,vijay2020measurementdrivenphase,fan2021selforganizederror,li2021conformal,ippoliti2021entanglementphasetransitions,ippoliti2022fractallogarithmicand,klocke2022topologicalorderand,lu2021spacetimeduality,ippoliti2021postselectionfreeentanglement,li2021robustdecodingin,li2021statisticalmechanicsmodel,li2021entanglementdomainwalls,li2021statisticalmechanicsof,feng2022measurementinducedphase,barratt2022transitions,zabalo2022operatorscalingdimensions,zabalo2020criticalpropertiesof,sierant2022universalbehaviorbeyond,iaconis2020measurementinducedphase,han2022measurementinducedcriticality,liu2022measurementinducedentanglement,sang2021entanglementnegativityat,shi2020entanglementnegativityat,weinstein2022measurementinducedpower,turkeshi2020measurementinducedcriticality,turkeshi2022measurementinducedcriticality,sierant2022measurementinducedphase,weinstein2022scramblngtransitionin,zabalo2022infiniterandomnesscriticality} with coinciding theoretical description in statistical mechanics~\cite{jian2021measurementinducedphase,lopezpiqueres2020meanfieldentanglement,vasseur2019entanglementtransitionsfrom,jian2020measurementinducedcriticality,nahum2921measurementandentanglement,nahum2023renormalizationgroupfor,bao2020theoryofthe,choi2020quantumerrorcorrection}.

On the other hand, the case of monitored free fermions is less understood. While the entanglement properties have been extensively discussed~\cite{cao2019entanglementina,alberton2021entanglementtransitionin,botzung2021engineereddissipationinduced,turkeshi2022enhancedentanglementnegativity,turkeshi2021measurementinducedentanglement,buchhold2022revealingmeasurementinduced,minato2022fateofmeasurementinduced,ladewig2022monitoredopenfermion,piccitto2022entanglementtransitionsin,gal2022volumetoarea,granet2022volumelawtoarealaw,muller2022measurementinduceddark,buchhold2021effectivetheoryfor,turkeshi2022entanglementtransitionsfrom,biella2021manybodyquantumzeno,turkeshi2022entanglementandcorrelation,nahum2020entanglementanddynamics}, purification has been only partially investigated. 
Ref.~\cite{fidkowski2021howdynamicalquantum} demonstrated that monitored fermionic systems are \emph{not} generic as they purify at most quadratically (and not exponentially) in system size. However, this derivation does not account for the interplay between symmetry and locality in quantum systems. 
Indeed, local unitary operations scramble less compared to the global ones in Ref.~\cite{fidkowski2021dynamical}, thus are expected to shorten the purification timescales. 

This manuscript tackles these issues by investigating the prominent role of symmetry on monitored free fermions. Concretely, we investigate $\mathbb{Z}_2$ parity preserving or U(1) number conserving circuits build of quadratic fermionic gates, the so-called \emph{Gaussian circuits}. We unveil the purification timescale studying the entropy of an ancilla initially entangled with the system~\cite{gullans2020scalableprobesof}. 
Irrespectively of the symmetry, frequent measurements drive the system in a \emph{pure phase}, with purification time logarithmic in system size $L$. Instead, the purification timescale in the \emph{mixed phase} at low measurement rates depends crucially on the system's symmetry. 
For $\mathbb{Z}_2$ symmetric circuits, we find $\tau_P\sim L \ln L$ supporting the recent analytical arguments based on the nonlinear sigma models (NLSM)~\cite{fava2023nonlinearsigmamodels}, cf. also Ref.~\cite{bao2021symmetryenrichedphases,chaoming2022criticalityandentanglement, chaoming2023measurementinducedentanglementtransitions,chaoming2022criticalityandentanglement}. Consequently, the mixed phase preserves a residual thermodynamic entropy at ${1\ll t \ll \tau_P} $~\cite{merritt2023entanglementtransitionswith}, enabling us to extract the correlation length critical exponent of the measurement-induced transition. 
Instead, U(1) conserving circuits purify in a time \emph{sublinear} in system size at any finite measurement rate. In the mixed phase, we find $\tau_P \sim L^{\alpha(p)}$ and continuously varying exponent $0\le \alpha(p)\lesssim 0.87$. As for the entanglement transitions~\cite{alberton2021entanglementtransitionin,buchhold2021effectivetheoryfor}, U(1) circuits reveal a Berezinskii-Kosterlitz-Thouless (BKT) measurement-induced transition separating the mixed and trivial phase.

\textit{Purification transition in Gaussian circuits.---} 
We consider Gaussian circuits of free-fermions with interspersed unitary-measurement discrete-time dynamics. 
Such a circuit is mappable to a random-tensor network~\cite{bao2021symmetryenrichedphases,chaoming2022criticalityandentanglement}, whose $N$-replica field theory is described by a non-linear sigma model
\begin{equation}
    \mathcal{S} =  \frac{1}{2g_B}\int dx dt \ \mathrm{tr}(\partial^\mu Q)^T \partial_\mu Q+\dots, \label{eq:nlsm}
\end{equation}
with $Q(x,t)$ a $N\times N$ matrix, $g_B$ the bare coupling fixed by timescales of the microscopic model, $N$ the number of replicas, and the ellipsis is on topological and symmetry-enlarged terms. 
We note that Eq.~\eqref{eq:nlsm} applies also to Anderson localizing models~\cite{bocquet2000disordered2dquasiparticles,senthil2000quasiparticlelocalizationin,chalker2001thermalmetalin,evers2008andersontransitions}, with the required replica limit ${N\to0}$ instead of ${N\to 1}$ as in the measurement problem. Nevertheless, this draws an analogy between entanglement in Gaussian circuits and conductance in disordered systems, with the running constant $g_R^{-1}(L)$ being a strength of entanglement (conductance) at a lengthscale $L$. 
Generic parity conserving Majorana circuits correspond to the DIII Altland-Zirnbauer symmetry class~\cite{chaoming2023measurementinducedentanglementtransitions,fava2023nonlinearsigmamodels}, with beta function $dg_R/d\ln \ell \simeq(N-2) g_R^2/(8\pi) + O(g_R^3)$ known perturbatively at any $N$. 
We see here the importance of the correct replica limit, as the relevance of the flow is fixed by $N$. 
In particular, $N<2$ implies a non-trivial stable phase governed by the $g_R\simeq 0 $ fixed point~\cite{fava2023nonlinearsigmamodels}, with the $N=2$ limit corresponding to BKT behavior cf. Ref.~\cite{bao2021symmetryenrichedphases}. 

The running coupling $g_R$ affects entanglement and purification, as well as the universality class of the MIPT. 
Consider first a pure state $|\Psi\rangle$ and a bipartition $A\cup B$. The R\'enyi entropy $S_n(\rho) \equiv -\ln\mathrm{tr}(\rho^n)$ of the reduced density matrix $\rho_A \equiv \mathrm{tr}_B |\Psi\rangle\langle \Psi|$ measures the entanglement between $A$ and $B$~\cite{amico2008entanglementinmanybody,Calabrese_2004,Calabrese_2009}. The running coupling $g_R^{-1}\simeq \ln L$ leads to the scaling in the entropy $S_n(\rho_A)\sim (n+1)/(96n) \ln^2 \ell_A$, with $\ell_A$ the length of $A$. 
The small prefactor hinders a precise entanglement scaling identification, which is susceptible to significant finite-size effects, resulting in conflicting numerical analysis~\cite{chaoming2023measurementinducedentanglementtransitions,fava2023nonlinearsigmamodels}. 
An equivalent picture emerges from studying the purification of an initially mixed state $\rho$, revealed by the evolution of the $S_2$ of a monitored system~\cite{gullans2020dynamicalpurificationphase,gullans2020scalableprobesof,merritt2023entanglementtransitionswith}. While rigorous arguments upper bound the purification timescale of monitored free fermions to be quadratic in system size~\cite{fidkowski2021howdynamicalquantum}, the renormalized constant affects the purification timescale as $\tau_P\sim L/g_R(\ln L)$. In particular, for the DIII class, the analytic prediction is $\tau_P \sim L \ln L$~\cite{fava2023nonlinearsigmamodels}. 
The advantage of considering the purification properties of the system is that they are extractable from a single ancilla qubit initially maximally entangled with the system~\cite{gullans2020scalableprobesof,block2022measurementinducedtransition,sharma2022measurementinducedcriticality}, leading to practical experimental proxies~\cite{koh2022experimentalrealizationof,noel2022measurementinducedquantum} and more robust numerical checks compared to those based on entanglement measures.
Including additional symmetries results in (potentially relevant) corrections to Eq.~\eqref{eq:nlsm}, altering $\tau_P$. 

\begin{figure}[t!]
    \centering
    \includegraphics[width=\columnwidth]{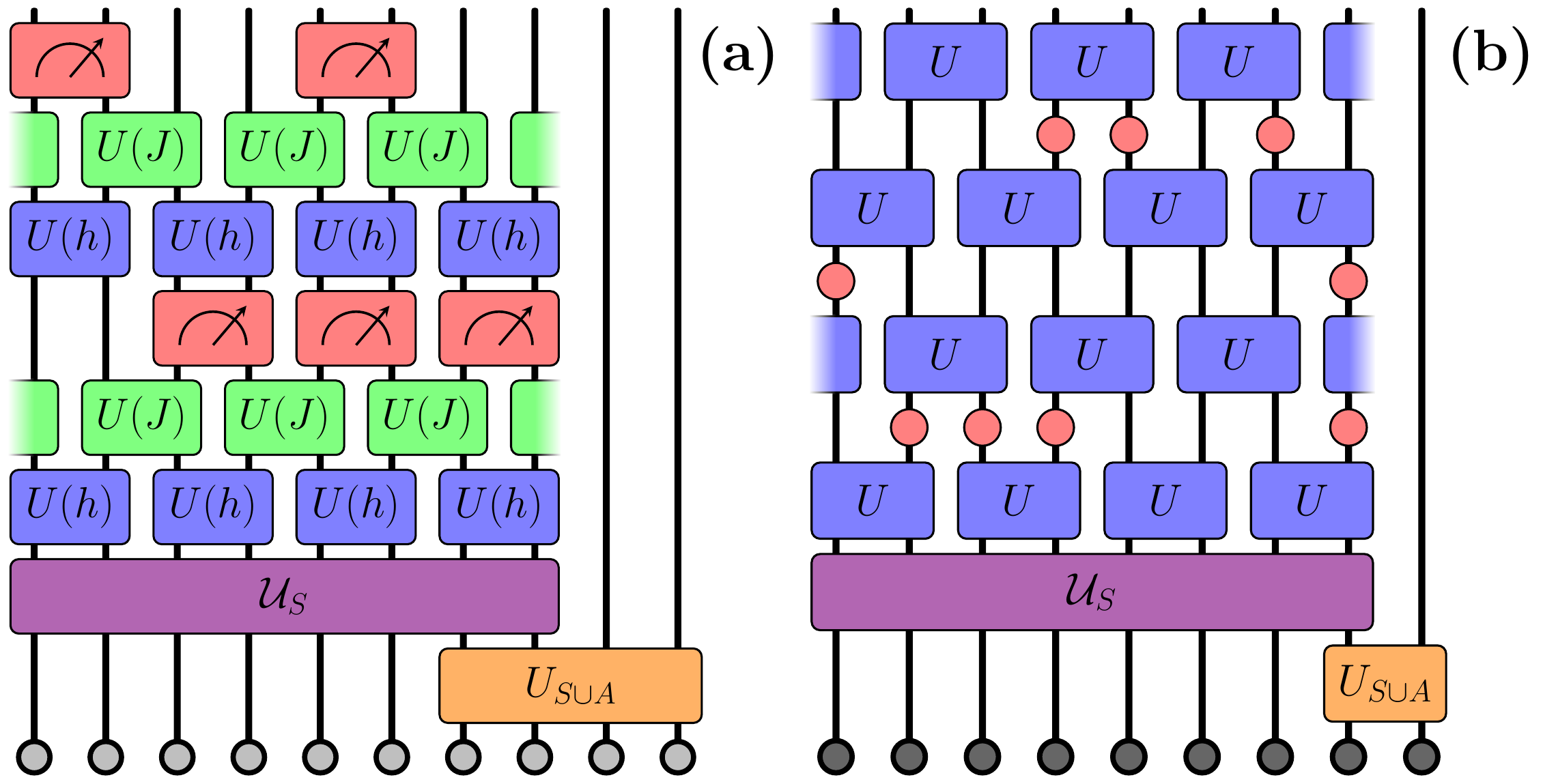}
    \caption{Cartoon of the Gaussian circuits of interest. (a) The Majorana model with $\mathbb{Z}_2$ with three types of gates: on-site Floquet unitaries $\hat{U}(g)$ (blue), nearest-neighboring $\hat{U}(J)$ gates (green), and on-site measurements with probability $p$ (red). (b) The stochastic U(1) conserving model with $\hat{U}$ random gates (blue) and onsite measurements (red), cf. text. The gate $\hat{U}_{S\cup A}$ act on system and ancillae, while $\hat{\mathcal{U}}_S$ on the state. Varying them and the initial state, we uncover different but physically equivalent facets of the problem. }
    \label{fig:cartoon}
\end{figure}

We elaborate on these points by investigating $\mathbb{Z}_2$ parity conserving and U(1) number conserving systems. In the former case, the mixed phase purification timescale matches the NLSM prediction~\cite{fava2023nonlinearsigmamodels}, indirectly supporting the flow $g_R\sim 1/\ln(L)$ for the DIII class circuits.
Instead, the purification timescale is always sublinear in system size for the U(1) case, with a BKT transition separating trivial and mixed phases. The latter exhibits ${\tau_P\sim L^{\alpha(p)}}$ with continuously varying exponent ${0<\alpha(p)\lesssim0.87}$, supporting a running constant $g_R\sim L^{1-\alpha(p)}$. 

\begin{figure*}[t!]
    \centering
    \includegraphics[width=\textwidth]{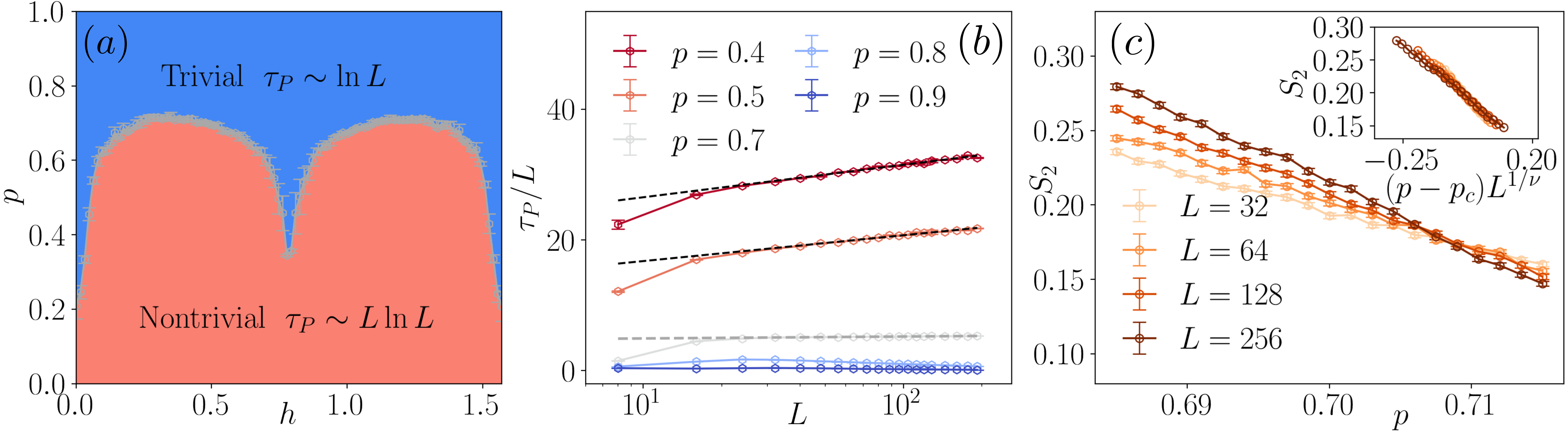}
    \caption{(a) Phase diagram of the Majorana circuit. For $p<p_c(h)$, the system is in a non-trivial mixed phase characterized by a purification timescale $\tau_P\sim L\ln L$ (pink region). Instead, for $p>p_c$, the system is in the purified phase, with $\tau  = \mathrm{const}$. The critical line $p_c(h)$ is determined numerically and has the limit of $\tau_P\sim L$. (b) Scaling of the typical purification timescale $\tau_P$ with system sizes for various measurement rates. The dashed lines are logarithmic fits $f(x) = a \ln x + b$, with $a,b$ two fitting parameters. The dashed grey line is a constant fit. (c) The residual entropy of an initially maximally mixed state at circuit depth $t=L$. Inset: data collapse with $p_c = 0.707$ and $\nu=2.1$. 
    } 
    \label{fig:ising}
\end{figure*}
For concreteness, we discuss implementations depicted in Fig.~\ref{fig:cartoon}. The first example is based on an Ising Trotterization, cf. Fig.~\ref{fig:cartoon}(a).
Each layer is given by quadratic operations of Majorana fermions $\{\hat{\gamma}_a,\hat{\gamma}_b\}=2\delta_{ab}$
\begin{equation}
    \hat{K}_p = \left[\prod_{j=1}^L \hat{M}_{2j-1,2j}^{r_j}\right] \left[\prod_{j=1}^L \hat{U}_{2j,2j+1}(J)\right]\prod_{j=1}^L\hat{U}_{2j-1,2j}(h)\label{eq:kising}
\end{equation}
where $r_j=1$ with probability $p$, otherwise $r_j=0$ and periodic boundary conditions apply. The probability $p$ is one of the parameters of the model controlling the rate of measurements. The unitaries in Eq.~\eqref{eq:kising} are fixed by $\hat{U}_{a,b}(\alpha) = \exp(- \alpha \hat\gamma_{a}\hat\gamma_{b})$ and the on-site measurements act as
\begin{equation}
    \hat{M}_{a,b}\rho\hat{M}_{a,b} = \frac{1}{p_\pm}\frac{1\pm  i \hat{\gamma}_{a} \hat{\gamma}_{b}}{2}\rho \frac{1\pm i \hat{\gamma}_{a} \hat{\gamma}_{b}}{2} 
\end{equation}
with the sign randomly chosen according to the Born rule probabilities $p_\pm = (1\pm \mathrm{tr}[i \hat{\gamma}_{a} \hat{\gamma}_{b} \rho])/2$~\cite{bravyi2004lagrangianrepresentationfor}. We infer the layer dependency on the measurement registry that fixed the quantum trajectory realization and call this system the Majorana circuit. The dynamics preserve the  $\mathbb{Z}_2$ parity symmetry of the system. 

Instead,  the U(1) preserving dynamics is realized with a stochastic circuit of Dirac fermions  $\{\hat{c}_a,\hat{c}_b^\dagger\}=\delta_{ab}$ with layers 
\begin{equation}
    \hat{K}_p^2 = \left(\prod_{j=1}^L \hat{M}_j^{r_j}\right)\prod_{i \in e} \hat{U}_{i,i+1}\left(\prod_{j=1}^L \hat{M}_j^{r_j}\right)\prod_{i \in o} \hat{U}_{i,i+1},
\end{equation}
with $i\in \textit{e/o}$ running over even/odd sites. Periodic boundary conditions are assumed and $r_j=1$ with probability $p$, otherwise $r_j=0$. The unitary gates are given by $\hat{U}_{a,b} = \exp[-2 i\beta (\hat{c}_{a}^\dagger\hat{c}_b+\mathrm{h.c.})]$ with $\beta$ a random number in $[0,\pi]$, while  $\hat{M}_{a}$ is an on-site measurement of the particle density $\hat{n}_a=\hat{c}^\dagger_a\hat{c}_a$ 
\begin{equation}
  \hat{M}_a\rho\hat{M}_a=\begin{cases} \displaystyle \frac{\hat{n}_a \rho \hat{n}_a}{\mathrm{tr}(\hat{n}_a\rho)},\quad \text{with probability } \mathrm{tr}(\hat{n}_a\rho)\\
    \displaystyle\frac{(1-\hat{n}_a) \rho (1- \hat{n}_a)}{1-\mathrm{tr}(\hat{n}_a\rho)},\ \text{otherwise.}
    \end{cases}\nonumber
\end{equation}
Again we infer the measurement registry dependence of $\hat{K}_p$ and conveniently denote this as the Dirac circuit. This dynamics has a U(1) symmetry generated by the total fermionic number $\hat{N} = \sum_k \hat{n}_k$~\cite{fidkowski2021howdynamicalquantum}. 

The models of interest are simulatable with just polynomial complexity in system size as, by Gaussianity, the whole evolution reflects in the dynamics of the correlation matrix of the system~\cite{bravyi2004lagrangianrepresentationfor,turkeshi2022enhancedentanglementnegativity}, respectively $M_{a,b}=\mathrm{tr}(\rho i [\hat{\gamma}_a,\hat{\gamma}_b]/2)$ ($a,b=1,\dots,2L$) and $C_{a,b} = \mathrm{tr}(\rho \hat{c}^\dagger_a \hat{c}_b) $ ($a,b=1,\dots,L$) for the Majorana and Dirac circuits~\cite{supmat}. 
We study a complementary but physically equivalent framework, based on the system circuit evolution $\rho_{S\cup A}(t) = \hat{K}^t_{p,S}{\rho_{S\cup A}(0)}(\hat{K}^t_{p,S})^\dagger$ of a system and ancillae initial state $\rho_{S\cup A}(0)=\hat{\mathcal{U}}_S \hat{U}_{S\cup A}\rho_{0,S}\otimes\rho_{0,A}\hat{U}_{S\cup A}^\dagger\hat{\mathcal{U}}_S^\dagger$, cf. Fig.~\ref{fig:cartoon}.  
The latter is obtained from the uncorrelated initial density matrices $\rho_{0,S}$ and $\rho_{0,A}$ applying $\hat{U}_{S\cup A}$ and $\hat{\mathcal{U}}_S$, respectively, a system-ancillae unitary operation and a system unitary gate. 
Specifically, we consider: (i) the purification of an ancilla system initially entangled to the system (with $\rho_{0,S/A}$ a product state, $\hat{U}_{S\cup A}$ a maximally entangling gate, and $\hat{\mathcal{U}}_S$ a scrambling gate), (ii) the purification of a maximally mixed state (with $\rho_{0,S} = \openone/2^L$ and $\hat{U}_{S\cup A}$, $\hat{\mathcal{U}}_S$ both identity gates, leaving the ancilla and system decoupled).
In both instances, the purification properties are revealed by the residual entropy $S_2(\rho_S(t))$. In particular, we extract the purification timescale  $\tau_P$ from the former protocol as the characteristic timescale at which $S_2(\rho_S(t))=0$~\footnote{Indeed, in this case, the initial system-ancilla state is pure, $S_2(\rho_S(t))=S_2(\rho_A(t))$ allowing considerable simplification in the numerical calculations.}.

\textit{Numerical results.---} 
We begin by numerically investigating the Majorana circuit, cf.~\ref{fig:cartoon}(a). The results are summarized in the phase diagram in Fig.~\ref{fig:ising}(a), where we fix $J=0.5$. 
As we detail below, we find the critical line $p=p_c(h)$ in grey separating a non-trivial mixed phase for $p<p_c(h)$ (in red in the figure) and a purifying phase for $p>p_c(h)$ (in blue) studying the residual entropy. 
Note that the phase diagram is periodic with period $\pi$ as a result of the choice of gates, cf. Eq.~\eqref{eq:kising}, and here we show only one branch. In particular, we expect $p_c=0$ as $h\to \pi$, since the $\hat{U}_{2j,2j-1}(h)$ become trivial. 

\begin{figure*}[t!]
    \centering
    \includegraphics[width=\textwidth]{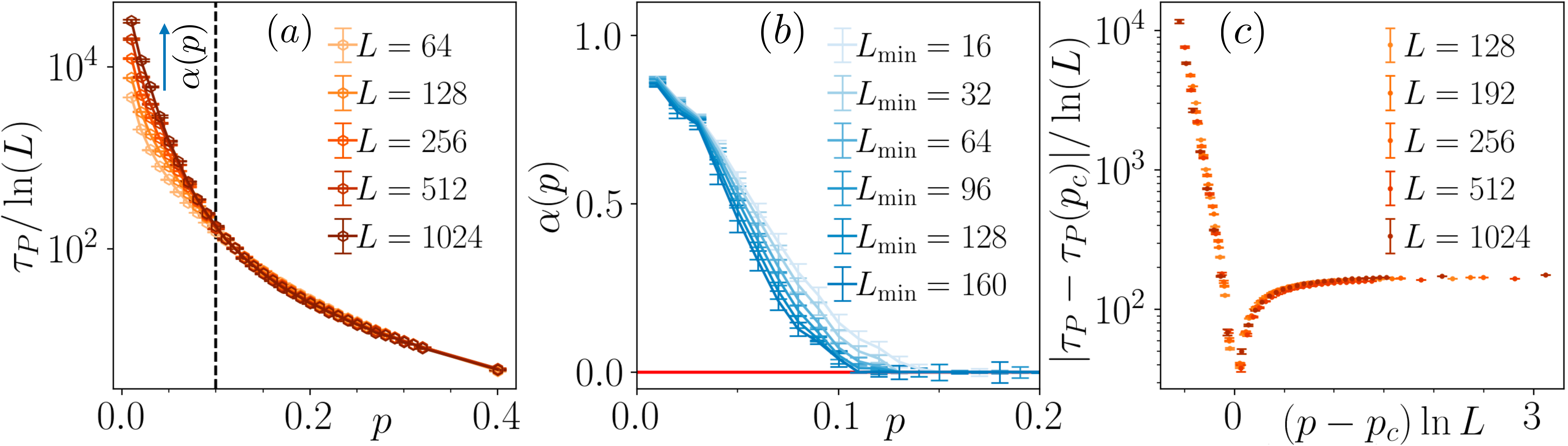}
    \caption{(a) Purification timescale $\tau_P/\ln(L)$ for the Dirac circuit with U(1)  for different system sizes  $L\le 1024$ and values of $p\in [0,0.4]$. The critical rate $p_c$ separates the mixed phase $p<p_c$, where one sees $\tau_P \sim L^{\alpha(p)}$ with $0<\alpha(p)\lesssim 0.87$, from a purified phase $p>p_c$ (b) The continuously varying exponent $\alpha(p)$ is obtained considering larger system sizes for the fit, cf. text. For $p>p_c$ it is compatible to zero, while for $p<p_c$ if grows approaching ${p\to0}$, saturating to a value $\sim 0.87$. (c) The data collapse for a BKT finite size scaling with the optimal $p_c= 0.10(1)$. We remark the cost function minimization involves only a single fitting parameter.   \label{fig:u1}}
\end{figure*}

We characterize the phases by first considering the purification of two Majorana ancillae, virtually located at the sites $L+1, L+2$ in Fig.~\ref{fig:ising}(a). We initialize $\rho^\mathrm{majo}_{S\cup A}(0)$ by locally entangling the ancillae with the system, and performing a scrambling operation on the system. The particular choice of scrambling operation does not change the qualitative physics, cf.~\cite{supmat}. 
We identify the purification timescale as $\tau_P=\mathrm{median}(t_P)$ and study its behavior for ${\mathcal{N}_\mathrm{real} > 10^3}$ realizations of the circuit~\footnote{\textit{En passant}, we note that considering other indicators such as the average value give qualitatively similar results.}. 
The results are reported in Fig.~\ref{fig:ising}(b) at $h=0.3$. 
In the mixed phase ($p<p_c\simeq 0.7$), we find $\tau_P\sim L\ln L$ as expected by the NLSM. 
Close to the critical value $p\approx p_c$, the system approaches a linear scaling $\tau_P\sim L$, while for $p>p_c$, the $\tau_P$ crossover to $\sim \log L$ scaling. This logarithmic growth has a trivial origin and can be understood since the number of unmeasured sites must become $O(1)$ at $t \sim \tau_P$, i.e.
$(1-p)^{\tau_P} L \lesssim 1 $.
The ancillae purification timescale, therefore, gives a clear separation between the non-trivial and mixed phases. This should be compared with the entanglement measures on extensive subsystems, where due to the significant finite-size effects, the identification of the scaling is more subtle (for instance, see discrepancies among Ref.~\cite{chaoming2023measurementinducedentanglementtransitions,fava2023nonlinearsigmamodels}). 

The superlinear $\tau_P$ in the mixed phase implies that the maximally mixed state $\rho_0 =\hat{\openone}/2^L$ will sustain a residual entropy after a circuit evolution of depth $t\sim O(L)$. For such depths, the residual entropy $S_2 = \overline{S_2(\hat{K}_p^t \rho_0 (\hat{K}_p^t)^\dagger)}$ distinguishes a mixed phase, where it attains a finite value, from a trivial phase, where it is exponentially suppressed in system size. 
As an example, we present the results for $t=L$ and $h=0.3$ in Fig.~\ref{fig:ising}(c). Compatible with the single qubit purification analysis, we observe a crossing behavior around $p\simeq 0.7$. 
This indicator estimates the $p_c$ with $\lesssim 2\%$ interval of confidence already with limited system sizes, cf. Fig.~\ref{fig:ising}(a) where we consider $t=L\le 64$ and averaging over $8000$ trajectories. 
Nevertheless, obtaining the critical properties requires a finite size scaling analysis, which we detail in~\cite{supmat}, and gives  $p_c^\star=0.707(3)$ and $\nu^\star=2.1(4)$. Our estimates, compatible with~\cite{merritt2023entanglementtransitionswith,chaoming2022criticalityandentanglement}, lead to the data collapse in Fig.~\ref{fig:ising}(c,inset). 
We see that the critical exponent is susceptible to more significant system size effects. We stress that an exact prediction for the exponent $\nu$ is not available even in the NLSM approach. The perturbative expansion around $N=2$, where BKT scaling implies ${\nu=\infty}$, can nonetheless give a possible justification of its numerically large value~\cite{fava2023nonlinearsigmamodels}.A conclusive answer based solely on the numerics is, therefore, out of reach for the considered system sizes and urges for more refined analytical insights.

We now investigate the purification of the Dirac circuit by considering the purification timescale of an ancilla maximally entangled with the system and virtually located at $L+1$, cf. Fig.~\ref{fig:u1}. 
Our numerical analysis is reported in Fig.~\ref{fig:u1}. First, we find that for any $p$, the purification timescale is \emph{sublinear}, cf. Fig.~\ref{fig:u1}(a). As for the Majorana circuit, for $p\ge p_c\simeq 0.1$ the system has a logarithmic purification timescale set by the scrambled initial state. In contrast, for $p\le p_c$, the scaling is power-law $L^{\alpha(p)}$, with an exponent $0< \alpha(p)<1$ which we extract from a fit~\footnote{To account for the logarithmic behavior for high rates $p\simeq 1$, we consider $f(x) = \beta (L^\alpha-1)/\alpha$. \textit{En passant}, we note that the available data cannot exclude the presence of multiplicative logarithms, arising in analogy to $\mathbb{Z}_2$ case, cf.~\cite{supmat}.}.
To account for finite-size effects, we sequentially include only $L\ge L_{\min}$, with ${L_{\min} = 16\div 160}$, cf. Fig.~\ref{fig:u1}(b). For $p>p_c$, the result is compatible with $\alpha(p)=0$, confirming the logarithmic system size scaling of $\tau_P$. 
On the other hand, for $p<p_c$, the power-law exponent continuously varies with the rate $p$, and for $p\to 0$, it attains values close to $\alpha(0)\simeq 0.87$. 
This preliminary analysis demonstrates a more subtle phenomenology compared to the Majorana circuit, suggesting a scaling theory beyond the power law diverging one of the NLSM. We, therefore, look for a BKT phase transition, where the correlation length diverges exponentially in  $1/(p-p_c)$. Specifically, we consider the scaling variable $x=(p-p_c)\ln L$ and, to account for the logarithmic corrections at $p=p_c$, we perform the finite size scaling on $\tau_P-\tau_P(p_c)$. In Fig.~\ref{fig:u1}(c), we show the excellent data collapse, particularly when considering that the fit is performed with only one free parameter, $p_c$.
These results supports the hypothesis that the $U(!)$
number conservation leads to a measurement-induced phase transition in the BKT universality class, both for the Dirac circuits analysed here and continuously monitored fermions ~\cite{alberton2021entanglementtransitionin,buchhold2021effectivetheoryfor}. 

\textit{Conclusion.---} 
We analyzed the purification dynamics of monitored fermionic circuits, highlighting the role of symmetry in determining the purification timescales. 
For systems with a $\mathbb{Z}_2$ parity conservation, our results provide robust support to recent analytical arguments~\cite{fava2023nonlinearsigmamodels}. In particular, we find a purification timescale that increases as $\tau_P\sim L\ln L $ in the mixed phase. 
We benchmarked these results considering the residual entropy at $t=L\ll \tau_P$, demonstrating a phase transition signalled by a crossing in the finite-size scaling of $S_2$ and extracting the correlation length critical exponent. 
For systems with a U(1) conservation law, the purification timescale is sublinear in system size for any measurement rate and supports the presence of BKT stemming from the extended symmetry. Within the framework of NLSM introduced in~\cite{fava2023nonlinearsigmamodels}, the available system sizes demonstrate this critical behavior separating a purifying phase from a mixed phase with $\tau_P\sim L^{\alpha(p)}$, indirectly supporting a running constant $g_R\sim L^{1-\alpha}$. 
We also note that the presence of a continuously varying exponent $0\le \alpha(p)\lesssim 0.87$ is compatible with previous studies on continuously monitored fermionic systems, as expected by virtue of universality.

In the continuous time limit, Ref.~\cite{buchhold2021effectivetheoryfor} found a BKT phase transition for U(1) conserving fermions for $N=2$ replica. Nevertheless, Ref.~\cite{fava2023nonlinearsigmamodels,bao2021symmetryenrichedphases,chaoming2023measurementinducedentanglementtransitions} demonstrate the importance of the replica limit $N\to 1$ to determine the measurement-induced universal properties of the system.  It is important and left for further investigation to analytically demonstrate the robustness of the BKT measurement-induced transition in the correct replica limit, as well as the rightful characterization of the mixed phase purification timescales. 
Moreover, in line with Ref.~\cite{bao2021symmetryenrichedphases,buccholdtoappear}, we expect that various discrete or continuous symmetries would enrich the phase diagram of the monitored system and leave hallmarks in the purification timescales. 
Lastly, our findings demonstrate the mixed phase of fermionic systems presents non-trivial structural properties that are revealable within the projective ensemble~\cite{ho2022exactemergentquantum,Claeys2022emergentquantum,Ippoliti2022solvablemodelofdeep,ippoliti2022dynamicalpurificationand,lucas2022generalizeddeepthermalization,lydzba2022}, for which the reduced density matrix reveals mean value aspects.
\begin{acknowledgments}
\textit{Acknowledgments.---} We thank M. Fava, A. Nahum, and L. Piroli for the discussion on their recent results. 
X.T. is indebted to B. Bauer, M. Buchhold, C.-M. Jian, A. W. W. Ludwig, H. Shapourian for discussion on related topics, and thanks R. Fazio, M. Schir\`o, P. Sierant for discussion and collaborations on related topics. He is supported by the ANR grant ``NonEQuMat.''
(ANR-19-CE47-0001) and computational resources on the Coll\'ge de France IPH cluster.  This work was granted access to the HPC resources of IDRIS under the allocation 2022-AD010513967,  made by GENCI. J.D.N. and H.L. are supported by the ERC Starting Grant 101042293 (HEPIQ). 
ADL acknowledges support by the ANR JCJC grant
ANR-21-CE47-0003 (TamEnt).
\end{acknowledgments}

\bibliography{biblio}

\begin{thebibliography}{140}%
\makeatletter
\providecommand \@ifxundefined [1]{%
 \@ifx{#1\undefined}
}%
\providecommand \@ifnum [1]{%
 \ifnum #1\expandafter \@firstoftwo
 \else \expandafter \@secondoftwo
 \fi
}%
\providecommand \@ifx [1]{%
 \ifx #1\expandafter \@firstoftwo
 \else \expandafter \@secondoftwo
 \fi
}%
\providecommand \natexlab [1]{#1}%
\providecommand \enquote  [1]{``#1''}%
\providecommand \bibnamefont  [1]{#1}%
\providecommand \bibfnamefont [1]{#1}%
\providecommand \citenamefont [1]{#1}%
\providecommand \href@noop [0]{\@secondoftwo}%
\providecommand \href [0]{\begingroup \@sanitize@url \@href}%
\providecommand \@href[1]{\@@startlink{#1}\@@href}%
\providecommand \@@href[1]{\endgroup#1\@@endlink}%
\providecommand \@sanitize@url [0]{\catcode `\\12\catcode `\$12\catcode
  `\&12\catcode `\#12\catcode `\^12\catcode `\_12\catcode `\%12\relax}%
\providecommand \@@startlink[1]{}%
\providecommand \@@endlink[0]{}%
\providecommand \url  [0]{\begingroup\@sanitize@url \@url }%
\providecommand \@url [1]{\endgroup\@href {#1}{\urlprefix }}%
\providecommand \urlprefix  [0]{URL }%
\providecommand \Eprint [0]{\href }%
\providecommand \doibase [0]{https://doi.org/}%
\providecommand \selectlanguage [0]{\@gobble}%
\providecommand \bibinfo  [0]{\@secondoftwo}%
\providecommand \bibfield  [0]{\@secondoftwo}%
\providecommand \translation [1]{[#1]}%
\providecommand \BibitemOpen [0]{}%
\providecommand \bibitemStop [0]{}%
\providecommand \bibitemNoStop [0]{.\EOS\space}%
\providecommand \EOS [0]{\spacefactor3000\relax}%
\providecommand \BibitemShut  [1]{\csname bibitem#1\endcsname}%
\let\auto@bib@innerbib\@empty
\bibitem [{\citenamefont {Georgescu}\ \emph {et~al.}(2014)\citenamefont
  {Georgescu}, \citenamefont {Ashhab},\ and\ \citenamefont
  {Nori}}]{georgescu2014quantumsimulation}%
  \BibitemOpen
  \bibfield  {author} {\bibinfo {author} {\bibfnamefont {I.~M.}\ \bibnamefont
  {Georgescu}}, \bibinfo {author} {\bibfnamefont {S.}~\bibnamefont {Ashhab}},\
  and\ \bibinfo {author} {\bibfnamefont {F.}~\bibnamefont {Nori}},\ }\href
  {https://doi.org/10.1103/RevModPhys.86.153} {\bibfield  {journal} {\bibinfo
  {journal} {Rev. Mod. Phys.}\ }\textbf {\bibinfo {volume} {86}},\ \bibinfo
  {pages} {153} (\bibinfo {year} {2014})}\BibitemShut {NoStop}%
\bibitem [{\citenamefont {Ferris}\ \emph {et~al.}(2022)\citenamefont {Ferris},
  \citenamefont {Rasmusson}, \citenamefont {Bronn},\ and\ \citenamefont
  {Lanes}}]{ferris2022quantumsimulationon}%
  \BibitemOpen
  \bibfield  {author} {\bibinfo {author} {\bibfnamefont {K.~J.}\ \bibnamefont
  {Ferris}}, \bibinfo {author} {\bibfnamefont {A.~J.}\ \bibnamefont
  {Rasmusson}}, \bibinfo {author} {\bibfnamefont {N.~T.}\ \bibnamefont
  {Bronn}},\ and\ \bibinfo {author} {\bibfnamefont {O.}~\bibnamefont {Lanes}},\
  }\href@noop {} {} (\bibinfo {year} {2022}),\ \Eprint
  {https://arxiv.org/abs/2209.02795} {arXiv:2209.02795 [quant-ph]} \BibitemShut
  {NoStop}%
\bibitem [{\citenamefont {Fraxanet}\ \emph {et~al.}(2022)\citenamefont
  {Fraxanet}, \citenamefont {Salamon},\ and\ \citenamefont
  {Lewenstein}}]{fraxanet2022comingdecasesof}%
  \BibitemOpen
  \bibfield  {author} {\bibinfo {author} {\bibfnamefont {J.}~\bibnamefont
  {Fraxanet}}, \bibinfo {author} {\bibfnamefont {T.}~\bibnamefont {Salamon}},\
  and\ \bibinfo {author} {\bibfnamefont {M.}~\bibnamefont {Lewenstein}},\
  }\href@noop {} {} (\bibinfo {year} {2022}),\ \Eprint
  {https://arxiv.org/abs/2204.08905} {arXiv:2204.08905 [quant-ph]} \BibitemShut
  {NoStop}%
\bibitem [{\citenamefont {Pezz\`e}\ \emph {et~al.}(2018)\citenamefont
  {Pezz\`e}, \citenamefont {Smerzi}, \citenamefont {Oberthaler}, \citenamefont
  {Schmied},\ and\ \citenamefont {Treutlein}}]{pezze2018quantummtrologywith}%
  \BibitemOpen
  \bibfield  {author} {\bibinfo {author} {\bibfnamefont {L.}~\bibnamefont
  {Pezz\`e}}, \bibinfo {author} {\bibfnamefont {A.}~\bibnamefont {Smerzi}},
  \bibinfo {author} {\bibfnamefont {M.~K.}\ \bibnamefont {Oberthaler}},
  \bibinfo {author} {\bibfnamefont {R.}~\bibnamefont {Schmied}},\ and\ \bibinfo
  {author} {\bibfnamefont {P.}~\bibnamefont {Treutlein}},\ }\href
  {https://doi.org/10.1103/RevModPhys.90.035005} {\bibfield  {journal}
  {\bibinfo  {journal} {Rev. Mod. Phys.}\ }\textbf {\bibinfo {volume} {90}},\
  \bibinfo {pages} {035005} (\bibinfo {year} {2018})}\BibitemShut {NoStop}%
\bibitem [{\citenamefont {Desaules}\ \emph {et~al.}(2022)\citenamefont
  {Desaules}, \citenamefont {Pietracaprina}, \citenamefont {Papi~\'{c}},
  \citenamefont {Goold},\ and\ \citenamefont
  {Pappalardi}}]{desaules2022extensivemultipartiteentanglement}%
  \BibitemOpen
  \bibfield  {author} {\bibinfo {author} {\bibfnamefont {J.-Y.}\ \bibnamefont
  {Desaules}}, \bibinfo {author} {\bibfnamefont {F.}~\bibnamefont
  {Pietracaprina}}, \bibinfo {author} {\bibfnamefont {Z.}~\bibnamefont
  {Papi~\'{c}}}, \bibinfo {author} {\bibfnamefont {J.}~\bibnamefont {Goold}},\
  and\ \bibinfo {author} {\bibfnamefont {S.}~\bibnamefont {Pappalardi}},\
  }\href {https://doi.org/10.1103/PhysRevLett.129.020601} {\bibfield  {journal}
  {\bibinfo  {journal} {Phys. Rev. Lett.}\ }\textbf {\bibinfo {volume} {129}},\
  \bibinfo {pages} {020601} (\bibinfo {year} {2022})}\BibitemShut {NoStop}%
\bibitem [{\citenamefont {Dooley}(2021)}]{dooley2021robustquantumsensing}%
  \BibitemOpen
  \bibfield  {author} {\bibinfo {author} {\bibfnamefont {S.}~\bibnamefont
  {Dooley}},\ }\href {https://doi.org/10.1103/PRXQuantum.2.020330} {\bibfield
  {journal} {\bibinfo  {journal} {PRX Quantum}\ }\textbf {\bibinfo {volume}
  {2}},\ \bibinfo {pages} {020330} (\bibinfo {year} {2021})}\BibitemShut
  {NoStop}%
\bibitem [{\citenamefont {Dooley}\ \emph {et~al.}(2018)\citenamefont {Dooley},
  \citenamefont {Hanks}, \citenamefont {Nakayama}, \citenamefont {Munro},\ and\
  \citenamefont {Nemoto}}]{dooley2018robustquantumsensing}%
  \BibitemOpen
  \bibfield  {author} {\bibinfo {author} {\bibfnamefont {S.}~\bibnamefont
  {Dooley}}, \bibinfo {author} {\bibfnamefont {M.}~\bibnamefont {Hanks}},
  \bibinfo {author} {\bibfnamefont {S.}~\bibnamefont {Nakayama}}, \bibinfo
  {author} {\bibfnamefont {W.~J.}\ \bibnamefont {Munro}},\ and\ \bibinfo
  {author} {\bibfnamefont {K.}~\bibnamefont {Nemoto}},\ }\bibfield  {journal}
  {\bibinfo  {journal} {npj Quantum Inf.}\ }\textbf {\bibinfo {volume} {4}},\
  \href {https://doi.org/10.1038/s41534-018-0073-3} {10.1038/s41534-018-0073-3}
  (\bibinfo {year} {2018})\BibitemShut {NoStop}%
\bibitem [{\citenamefont {Dooley}\ \emph {et~al.}(2023)\citenamefont {Dooley},
  \citenamefont {Pappalardi},\ and\ \citenamefont
  {Goold}}]{dolay2023entanglementenhancedmetrology}%
  \BibitemOpen
  \bibfield  {author} {\bibinfo {author} {\bibfnamefont {S.}~\bibnamefont
  {Dooley}}, \bibinfo {author} {\bibfnamefont {S.}~\bibnamefont {Pappalardi}},\
  and\ \bibinfo {author} {\bibfnamefont {J.}~\bibnamefont {Goold}},\ }\href
  {https://doi.org/10.1103/PhysRevB.107.035123} {\bibfield  {journal} {\bibinfo
   {journal} {Phys. Rev. B}\ }\textbf {\bibinfo {volume} {107}},\ \bibinfo
  {pages} {035123} (\bibinfo {year} {2023})}\BibitemShut {NoStop}%
\bibitem [{\citenamefont {Preskill}(2018)}]{preskill2018quantumcomputingin}%
  \BibitemOpen
  \bibfield  {author} {\bibinfo {author} {\bibfnamefont {J.}~\bibnamefont
  {Preskill}},\ }\href {https://doi.org/10.22331/q-2018-08-06-79} {\bibfield
  {journal} {\bibinfo  {journal} {{Quantum}}\ }\textbf {\bibinfo {volume}
  {2}},\ \bibinfo {pages} {79} (\bibinfo {year} {2018})}\BibitemShut {NoStop}%
\bibitem [{\citenamefont
  {Deutsch}(1991)}]{deutsch1991quantumstatisticalmechanics}%
  \BibitemOpen
  \bibfield  {author} {\bibinfo {author} {\bibfnamefont {J.~M.}\ \bibnamefont
  {Deutsch}},\ }\href {https://doi.org/10.1103/PhysRevA.43.2046} {\bibfield
  {journal} {\bibinfo  {journal} {Phys. Rev. A}\ }\textbf {\bibinfo {volume}
  {43}},\ \bibinfo {pages} {2046} (\bibinfo {year} {1991})}\BibitemShut
  {NoStop}%
\bibitem [{\citenamefont {Srednicki}(1994)}]{srednicki1994chaosandquantum}%
  \BibitemOpen
  \bibfield  {author} {\bibinfo {author} {\bibfnamefont {M.}~\bibnamefont
  {Srednicki}},\ }\href {https://doi.org/10.1103/PhysRevE.50.888} {\bibfield
  {journal} {\bibinfo  {journal} {Phys. Rev. E}\ }\textbf {\bibinfo {volume}
  {50}},\ \bibinfo {pages} {888} (\bibinfo {year} {1994})}\BibitemShut
  {NoStop}%
\bibitem [{\citenamefont {Srednicki}(1999)}]{Srednicki1999theapproachto}%
  \BibitemOpen
  \bibfield  {author} {\bibinfo {author} {\bibfnamefont {M.}~\bibnamefont
  {Srednicki}},\ }\href {https://doi.org/10.1088/0305-4470/32/7/007} {\bibfield
   {journal} {\bibinfo  {journal} {J. Phys. A}\ }\textbf {\bibinfo {volume}
  {32}},\ \bibinfo {pages} {1163} (\bibinfo {year} {1999})}\BibitemShut
  {NoStop}%
\bibitem [{\citenamefont {Rigol}\ \emph {et~al.}(2008)\citenamefont {Rigol},
  \citenamefont {Dunjko},\ and\ \citenamefont
  {Olshanii}}]{Rigol2008thermalizationandits}%
  \BibitemOpen
  \bibfield  {author} {\bibinfo {author} {\bibfnamefont {M.}~\bibnamefont
  {Rigol}}, \bibinfo {author} {\bibfnamefont {V.}~\bibnamefont {Dunjko}},\ and\
  \bibinfo {author} {\bibfnamefont {M.}~\bibnamefont {Olshanii}},\ }\href
  {https://doi.org/10.1038/nature06838} {\bibfield  {journal} {\bibinfo
  {journal} {Nature}\ }\textbf {\bibinfo {volume} {452}},\ \bibinfo {pages}
  {854} (\bibinfo {year} {2008})}\BibitemShut {NoStop}%
\bibitem [{\citenamefont {Polkovnikov}\ \emph {et~al.}(2011)\citenamefont
  {Polkovnikov}, \citenamefont {Sengupta}, \citenamefont {Silva},\ and\
  \citenamefont {Vengalattore}}]{polkovnikov2011nonequilibriumdynamicsof}%
  \BibitemOpen
  \bibfield  {author} {\bibinfo {author} {\bibfnamefont {A.}~\bibnamefont
  {Polkovnikov}}, \bibinfo {author} {\bibfnamefont {K.}~\bibnamefont
  {Sengupta}}, \bibinfo {author} {\bibfnamefont {A.}~\bibnamefont {Silva}},\
  and\ \bibinfo {author} {\bibfnamefont {M.}~\bibnamefont {Vengalattore}},\
  }\href {https://doi.org/10.1103/RevModPhys.83.863} {\bibfield  {journal}
  {\bibinfo  {journal} {Rev. Mod. Phys.}\ }\textbf {\bibinfo {volume} {83}},\
  \bibinfo {pages} {863} (\bibinfo {year} {2011})}\BibitemShut {NoStop}%
\bibitem [{\citenamefont {Foini}\ and\ \citenamefont
  {Kurchan}(2019)}]{foini2019eigenstatethermalizationhypothesis}%
  \BibitemOpen
  \bibfield  {author} {\bibinfo {author} {\bibfnamefont {L.}~\bibnamefont
  {Foini}}\ and\ \bibinfo {author} {\bibfnamefont {J.}~\bibnamefont
  {Kurchan}},\ }\href {https://doi.org/10.1103/PhysRevE.99.042139} {\bibfield
  {journal} {\bibinfo  {journal} {Phys. Rev. E}\ }\textbf {\bibinfo {volume}
  {99}},\ \bibinfo {pages} {042139} (\bibinfo {year} {2019})}\BibitemShut
  {NoStop}%
\bibitem [{\citenamefont {Pappalardi}\ \emph {et~al.}(2022)\citenamefont
  {Pappalardi}, \citenamefont {Foini},\ and\ \citenamefont
  {Kurchan}}]{pappalardi2022eigenstatethermalizationhypothesis}%
  \BibitemOpen
  \bibfield  {author} {\bibinfo {author} {\bibfnamefont {S.}~\bibnamefont
  {Pappalardi}}, \bibinfo {author} {\bibfnamefont {L.}~\bibnamefont {Foini}},\
  and\ \bibinfo {author} {\bibfnamefont {J.}~\bibnamefont {Kurchan}},\ }\href
  {https://doi.org/10.1103/PhysRevLett.129.170603} {\bibfield  {journal}
  {\bibinfo  {journal} {Phys. Rev. Lett.}\ }\textbf {\bibinfo {volume} {129}},\
  \bibinfo {pages} {170603} (\bibinfo {year} {2022})}\BibitemShut {NoStop}%
\bibitem [{\citenamefont {Fisher}\ \emph {et~al.}()\citenamefont {Fisher},
  \citenamefont {Khemani}, \citenamefont {Nahum},\ and\ \citenamefont
  {Vijay}}]{fisher2022randomquantumcircuits}%
  \BibitemOpen
  \bibfield  {author} {\bibinfo {author} {\bibfnamefont {M.~P.~A.}\
  \bibnamefont {Fisher}}, \bibinfo {author} {\bibfnamefont {V.}~\bibnamefont
  {Khemani}}, \bibinfo {author} {\bibfnamefont {A.}~\bibnamefont {Nahum}},\
  and\ \bibinfo {author} {\bibfnamefont {S.}~\bibnamefont {Vijay}},\
  }\href@noop {} {}\Eprint {https://arxiv.org/abs/2207.14280}
  {arXiv:2207.14280} \BibitemShut {NoStop}%
\bibitem [{\citenamefont {Gullans}\ and\ \citenamefont
  {Huse}(2020{\natexlab{a}})}]{gullans2020dynamicalpurificationphase}%
  \BibitemOpen
  \bibfield  {author} {\bibinfo {author} {\bibfnamefont {M.~J.}\ \bibnamefont
  {Gullans}}\ and\ \bibinfo {author} {\bibfnamefont {D.~A.}\ \bibnamefont
  {Huse}},\ }\href {https://doi.org/10.1103/PhysRevX.10.041020} {\bibfield
  {journal} {\bibinfo  {journal} {Phys. Rev. X}\ }\textbf {\bibinfo {volume}
  {10}},\ \bibinfo {pages} {041020} (\bibinfo {year}
  {2020}{\natexlab{a}})}\BibitemShut {NoStop}%
\bibitem [{\citenamefont {Gullans}\ and\ \citenamefont
  {Huse}(2020{\natexlab{b}})}]{gullans2020scalableprobesof}%
  \BibitemOpen
  \bibfield  {author} {\bibinfo {author} {\bibfnamefont {M.~J.}\ \bibnamefont
  {Gullans}}\ and\ \bibinfo {author} {\bibfnamefont {D.~A.}\ \bibnamefont
  {Huse}},\ }\href {https://doi.org/10.1103/PhysRevLett.125.070606} {\bibfield
  {journal} {\bibinfo  {journal} {Phys. Rev. Lett.}\ }\textbf {\bibinfo
  {volume} {125}},\ \bibinfo {pages} {070606} (\bibinfo {year}
  {2020}{\natexlab{b}})}\BibitemShut {NoStop}%
\bibitem [{\citenamefont {Lunt}\ \emph {et~al.}(2021)\citenamefont {Lunt},
  \citenamefont {Szyniszewski},\ and\ \citenamefont
  {Pal}}]{lunt2021measurementinducedcriticality}%
  \BibitemOpen
  \bibfield  {author} {\bibinfo {author} {\bibfnamefont {O.}~\bibnamefont
  {Lunt}}, \bibinfo {author} {\bibfnamefont {M.}~\bibnamefont {Szyniszewski}},\
  and\ \bibinfo {author} {\bibfnamefont {A.}~\bibnamefont {Pal}},\ }\href
  {https://doi.org/10.1103/PhysRevB.104.155111} {\bibfield  {journal} {\bibinfo
   {journal} {Phys. Rev. B}\ }\textbf {\bibinfo {volume} {104}},\ \bibinfo
  {pages} {155111} (\bibinfo {year} {2021})}\BibitemShut {NoStop}%
\bibitem [{\citenamefont {Sierant}\ \emph
  {et~al.}(2022{\natexlab{a}})\citenamefont {Sierant}, \citenamefont
  {Chiriac{\`{o}}}, \citenamefont {Surace}, \citenamefont {Sharma},
  \citenamefont {Turkeshi}, \citenamefont {Dalmonte}, \citenamefont {Fazio},\
  and\ \citenamefont {Pagano}}]{sierant2022dissipativefloquet}%
  \BibitemOpen
  \bibfield  {author} {\bibinfo {author} {\bibfnamefont {P.}~\bibnamefont
  {Sierant}}, \bibinfo {author} {\bibfnamefont {G.}~\bibnamefont
  {Chiriac{\`{o}}}}, \bibinfo {author} {\bibfnamefont {F.~M.}\ \bibnamefont
  {Surace}}, \bibinfo {author} {\bibfnamefont {S.}~\bibnamefont {Sharma}},
  \bibinfo {author} {\bibfnamefont {X.}~\bibnamefont {Turkeshi}}, \bibinfo
  {author} {\bibfnamefont {M.}~\bibnamefont {Dalmonte}}, \bibinfo {author}
  {\bibfnamefont {R.}~\bibnamefont {Fazio}},\ and\ \bibinfo {author}
  {\bibfnamefont {G.}~\bibnamefont {Pagano}},\ }\href
  {https://doi.org/10.22331/q-2022-02-02-638} {\bibfield  {journal} {\bibinfo
  {journal} {{Quantum}}\ }\textbf {\bibinfo {volume} {6}},\ \bibinfo {pages}
  {638} (\bibinfo {year} {2022}{\natexlab{a}})}\BibitemShut {NoStop}%
\bibitem [{\citenamefont {Zabalo}\ \emph {et~al.}(2020)\citenamefont {Zabalo},
  \citenamefont {Gullans}, \citenamefont {Wilson}, \citenamefont
  {Gopalakrishnan}, \citenamefont {Huse},\ and\ \citenamefont
  {Pixley}}]{zabalo2020criticalpropertiesof}%
  \BibitemOpen
  \bibfield  {author} {\bibinfo {author} {\bibfnamefont {A.}~\bibnamefont
  {Zabalo}}, \bibinfo {author} {\bibfnamefont {M.~J.}\ \bibnamefont {Gullans}},
  \bibinfo {author} {\bibfnamefont {J.~H.}\ \bibnamefont {Wilson}}, \bibinfo
  {author} {\bibfnamefont {S.}~\bibnamefont {Gopalakrishnan}}, \bibinfo
  {author} {\bibfnamefont {D.~A.}\ \bibnamefont {Huse}},\ and\ \bibinfo
  {author} {\bibfnamefont {J.~H.}\ \bibnamefont {Pixley}},\ }\href
  {https://doi.org/10.1103/PhysRevB.101.060301} {\bibfield  {journal} {\bibinfo
   {journal} {Phys. Rev. B}\ }\textbf {\bibinfo {volume} {101}},\ \bibinfo
  {pages} {060301} (\bibinfo {year} {2020})}\BibitemShut {NoStop}%
\bibitem [{\citenamefont {Sierant}\ \emph
  {et~al.}(2022{\natexlab{b}})\citenamefont {Sierant}, \citenamefont
  {Schir\`o}, \citenamefont {Lewenstein},\ and\ \citenamefont
  {Turkeshi}}]{sierant2022measurementinducedphase}%
  \BibitemOpen
  \bibfield  {author} {\bibinfo {author} {\bibfnamefont {P.}~\bibnamefont
  {Sierant}}, \bibinfo {author} {\bibfnamefont {M.}~\bibnamefont {Schir\`o}},
  \bibinfo {author} {\bibfnamefont {M.}~\bibnamefont {Lewenstein}},\ and\
  \bibinfo {author} {\bibfnamefont {X.}~\bibnamefont {Turkeshi}},\ }\href
  {https://doi.org/10.1103/PhysRevB.106.214316} {\bibfield  {journal} {\bibinfo
   {journal} {Phys. Rev. B}\ }\textbf {\bibinfo {volume} {106}},\ \bibinfo
  {pages} {214316} (\bibinfo {year} {2022}{\natexlab{b}})}\BibitemShut
  {NoStop}%
\bibitem [{\citenamefont {Lunt}\ \emph {et~al.}(2022)\citenamefont {Lunt},
  \citenamefont {Richter},\ and\ \citenamefont
  {Pal}}]{lunt2022quantumsimulationusing}%
  \BibitemOpen
  \bibfield  {author} {\bibinfo {author} {\bibfnamefont {O.}~\bibnamefont
  {Lunt}}, \bibinfo {author} {\bibfnamefont {J.}~\bibnamefont {Richter}},\ and\
  \bibinfo {author} {\bibfnamefont {A.}~\bibnamefont {Pal}},\ }\bibinfo {title}
  {Quantum simulation using noisy unitary circuits and measurements},\ in\
  \href {https://doi.org/10.1007/978-3-031-03998-0_10} {\emph {\bibinfo
  {booktitle} {Entanglement in Spin Chains: From Theory to Quantum Technology
  Applications}}},\ \bibinfo {editor} {edited by\ \bibinfo {editor}
  {\bibfnamefont {A.}~\bibnamefont {Bayat}}, \bibinfo {editor} {\bibfnamefont
  {S.}~\bibnamefont {Bose}},\ and\ \bibinfo {editor} {\bibfnamefont
  {H.}~\bibnamefont {Johannesson}}}\ (\bibinfo  {publisher} {Springer},\
  \bibinfo {address} {Cham},\ \bibinfo {year} {2022})\ pp.\ \bibinfo {pages}
  {251--284}\BibitemShut {NoStop}%
\bibitem [{\citenamefont {Potter}\ and\ \citenamefont
  {Vasseur}(2022)}]{potter2022entanglementdynamicsin}%
  \BibitemOpen
  \bibfield  {author} {\bibinfo {author} {\bibfnamefont {A.~C.}\ \bibnamefont
  {Potter}}\ and\ \bibinfo {author} {\bibfnamefont {R.}~\bibnamefont
  {Vasseur}},\ }\bibinfo {title} {Entanglement dynamics in hybrid quantum
  circuits},\ in\ \href {https://doi.org/10.1007/978-3-031-03998-0_9} {\emph
  {\bibinfo {booktitle} {Entanglement in Spin Chains: From Theory to Quantum
  Technology Applications}}},\ \bibinfo {editor} {edited by\ \bibinfo {editor}
  {\bibfnamefont {A.}~\bibnamefont {Bayat}}, \bibinfo {editor} {\bibfnamefont
  {S.}~\bibnamefont {Bose}},\ and\ \bibinfo {editor} {\bibfnamefont
  {H.}~\bibnamefont {Johannesson}}}\ (\bibinfo  {publisher} {Springer},\
  \bibinfo {address} {Cham},\ \bibinfo {year} {2022})\ pp.\ \bibinfo {pages}
  {211--249}\BibitemShut {NoStop}%
\bibitem [{\citenamefont {Rossini}\ and\ \citenamefont
  {Vicari}(2021)}]{rossini2021coherentanddissipative}%
  \BibitemOpen
  \bibfield  {author} {\bibinfo {author} {\bibfnamefont {D.}~\bibnamefont
  {Rossini}}\ and\ \bibinfo {author} {\bibfnamefont {E.}~\bibnamefont
  {Vicari}},\ }\href
  {https://doi.org/https://doi.org/10.1016/j.physrep.2021.08.003} {\bibfield
  {journal} {\bibinfo  {journal} {Phys. Rep.}\ }\textbf {\bibinfo {volume}
  {936}},\ \bibinfo {pages} {1} (\bibinfo {year} {2021})}\BibitemShut {NoStop}%
\bibitem [{\citenamefont {Chan}\ \emph {et~al.}(2019)\citenamefont {Chan},
  \citenamefont {Nandkishore}, \citenamefont {Pretko},\ and\ \citenamefont
  {Smith}}]{chan2019unitaryprojective}%
  \BibitemOpen
  \bibfield  {author} {\bibinfo {author} {\bibfnamefont {A.}~\bibnamefont
  {Chan}}, \bibinfo {author} {\bibfnamefont {R.~M.}\ \bibnamefont
  {Nandkishore}}, \bibinfo {author} {\bibfnamefont {M.}~\bibnamefont
  {Pretko}},\ and\ \bibinfo {author} {\bibfnamefont {G.}~\bibnamefont
  {Smith}},\ }\href {https://doi.org/10.1103/PhysRevB.99.224307} {\bibfield
  {journal} {\bibinfo  {journal} {Phys. Rev. B}\ }\textbf {\bibinfo {volume}
  {99}},\ \bibinfo {pages} {224307} (\bibinfo {year} {2019})}\BibitemShut
  {NoStop}%
\bibitem [{\citenamefont {Li}\ \emph {et~al.}(2019)\citenamefont {Li},
  \citenamefont {Chen},\ and\ \citenamefont
  {Fisher}}]{li2019measurementdrivenentanglement}%
  \BibitemOpen
  \bibfield  {author} {\bibinfo {author} {\bibfnamefont {Y.}~\bibnamefont
  {Li}}, \bibinfo {author} {\bibfnamefont {X.}~\bibnamefont {Chen}},\ and\
  \bibinfo {author} {\bibfnamefont {M.~P.~A.}\ \bibnamefont {Fisher}},\ }\href
  {https://doi.org/10.1103/PhysRevB.100.134306} {\bibfield  {journal} {\bibinfo
   {journal} {Phys. Rev. B}\ }\textbf {\bibinfo {volume} {100}},\ \bibinfo
  {pages} {134306} (\bibinfo {year} {2019})}\BibitemShut {NoStop}%
\bibitem [{\citenamefont {Skinner}\ \emph {et~al.}(2019)\citenamefont
  {Skinner}, \citenamefont {Ruhman},\ and\ \citenamefont
  {Nahum}}]{skinner2019measurementinducedphase}%
  \BibitemOpen
  \bibfield  {author} {\bibinfo {author} {\bibfnamefont {B.}~\bibnamefont
  {Skinner}}, \bibinfo {author} {\bibfnamefont {J.}~\bibnamefont {Ruhman}},\
  and\ \bibinfo {author} {\bibfnamefont {A.}~\bibnamefont {Nahum}},\ }\href
  {https://doi.org/10.1103/PhysRevX.9.031009} {\bibfield  {journal} {\bibinfo
  {journal} {Phys. Rev. X}\ }\textbf {\bibinfo {volume} {9}},\ \bibinfo {pages}
  {031009} (\bibinfo {year} {2019})}\BibitemShut {NoStop}%
\bibitem [{\citenamefont {Czischek}\ \emph {et~al.}(2021)\citenamefont
  {Czischek}, \citenamefont {Torlai}, \citenamefont {Ray}, \citenamefont
  {Islam},\ and\ \citenamefont {Melko}}]{czischek2021simulating}%
  \BibitemOpen
  \bibfield  {author} {\bibinfo {author} {\bibfnamefont {S.}~\bibnamefont
  {Czischek}}, \bibinfo {author} {\bibfnamefont {G.}~\bibnamefont {Torlai}},
  \bibinfo {author} {\bibfnamefont {S.}~\bibnamefont {Ray}}, \bibinfo {author}
  {\bibfnamefont {R.}~\bibnamefont {Islam}},\ and\ \bibinfo {author}
  {\bibfnamefont {R.~G.}\ \bibnamefont {Melko}},\ }\href
  {https://doi.org/10.1103/PhysRevA.104.062405} {\bibfield  {journal} {\bibinfo
   {journal} {Phys. Rev. A}\ }\textbf {\bibinfo {volume} {104}},\ \bibinfo
  {pages} {062405} (\bibinfo {year} {2021})}\BibitemShut {NoStop}%
\bibitem [{\citenamefont {Han}\ and\ \citenamefont
  {Chen}(2023)}]{han2022entanglementstructure}%
  \BibitemOpen
  \bibfield  {author} {\bibinfo {author} {\bibfnamefont {Y.}~\bibnamefont
  {Han}}\ and\ \bibinfo {author} {\bibfnamefont {X.}~\bibnamefont {Chen}},\
  }\href {https://doi.org/10.1103/PhysRevB.107.014306} {\bibfield  {journal}
  {\bibinfo  {journal} {Phys. Rev. B}\ }\textbf {\bibinfo {volume} {107}},\
  \bibinfo {pages} {014306} (\bibinfo {year} {2023})}\BibitemShut {NoStop}%
\bibitem [{\citenamefont {Minoguchi}\ \emph {et~al.}(2022)\citenamefont
  {Minoguchi}, \citenamefont {Rabl},\ and\ \citenamefont
  {Buchhold}}]{minoguchi2022continuousgaussianmeasurements}%
  \BibitemOpen
  \bibfield  {author} {\bibinfo {author} {\bibfnamefont {Y.}~\bibnamefont
  {Minoguchi}}, \bibinfo {author} {\bibfnamefont {P.}~\bibnamefont {Rabl}},\
  and\ \bibinfo {author} {\bibfnamefont {M.}~\bibnamefont {Buchhold}},\ }\href
  {https://doi.org/10.21468/SciPostPhys.12.1.009} {\bibfield  {journal}
  {\bibinfo  {journal} {SciPost Phys.}\ }\textbf {\bibinfo {volume} {12}},\
  \bibinfo {pages} {009} (\bibinfo {year} {2022})}\BibitemShut {NoStop}%
\bibitem [{\citenamefont {Altland}\ \emph {et~al.}(2022)\citenamefont
  {Altland}, \citenamefont {Buchhold}, \citenamefont {Diehl},\ and\
  \citenamefont {Micklitz}}]{altland2022dynamicsofmeasured}%
  \BibitemOpen
  \bibfield  {author} {\bibinfo {author} {\bibfnamefont {A.}~\bibnamefont
  {Altland}}, \bibinfo {author} {\bibfnamefont {M.}~\bibnamefont {Buchhold}},
  \bibinfo {author} {\bibfnamefont {S.}~\bibnamefont {Diehl}},\ and\ \bibinfo
  {author} {\bibfnamefont {T.}~\bibnamefont {Micklitz}},\ }\href
  {https://doi.org/10.1103/PhysRevResearch.4.L022066} {\bibfield  {journal}
  {\bibinfo  {journal} {Phys. Rev. Res.}\ }\textbf {\bibinfo {volume} {4}},\
  \bibinfo {pages} {L022066} (\bibinfo {year} {2022})}\BibitemShut {NoStop}%
\bibitem [{\citenamefont {Fuji}\ and\ \citenamefont
  {Ashida}(2020)}]{fuji2020measurementinducedquantum}%
  \BibitemOpen
  \bibfield  {author} {\bibinfo {author} {\bibfnamefont {Y.}~\bibnamefont
  {Fuji}}\ and\ \bibinfo {author} {\bibfnamefont {Y.}~\bibnamefont {Ashida}},\
  }\href {https://doi.org/10.1103/PhysRevB.102.054302} {\bibfield  {journal}
  {\bibinfo  {journal} {Phys. Rev. B}\ }\textbf {\bibinfo {volume} {102}},\
  \bibinfo {pages} {054302} (\bibinfo {year} {2020})}\BibitemShut {NoStop}%
\bibitem [{\citenamefont {Jian}\ \emph
  {et~al.}(2021{\natexlab{a}})\citenamefont {Jian}, \citenamefont {Yang},
  \citenamefont {Bi},\ and\ \citenamefont {Chen}}]{jian2021yangleeedge}%
  \BibitemOpen
  \bibfield  {author} {\bibinfo {author} {\bibfnamefont {S.-K.}\ \bibnamefont
  {Jian}}, \bibinfo {author} {\bibfnamefont {Z.-C.}\ \bibnamefont {Yang}},
  \bibinfo {author} {\bibfnamefont {Z.}~\bibnamefont {Bi}},\ and\ \bibinfo
  {author} {\bibfnamefont {X.}~\bibnamefont {Chen}},\ }\href
  {https://doi.org/10.1103/PhysRevB.104.L161107} {\bibfield  {journal}
  {\bibinfo  {journal} {Phys. Rev. B}\ }\textbf {\bibinfo {volume} {104}},\
  \bibinfo {pages} {L161107} (\bibinfo {year}
  {2021}{\natexlab{a}})}\BibitemShut {NoStop}%
\bibitem [{\citenamefont {Jin}\ and\ \citenamefont
  {Martin}(2022)}]{jin2022KPZ}%
  \BibitemOpen
  \bibfield  {author} {\bibinfo {author} {\bibfnamefont {T.}~\bibnamefont
  {Jin}}\ and\ \bibinfo {author} {\bibfnamefont {D.~G.}\ \bibnamefont
  {Martin}},\ }\href {https://doi.org/10.1103/PhysRevLett.129.260603}
  {\bibfield  {journal} {\bibinfo  {journal} {Phys. Rev. Lett.}\ }\textbf
  {\bibinfo {volume} {129}},\ \bibinfo {pages} {260603} (\bibinfo {year}
  {2022})}\BibitemShut {NoStop}%
\bibitem [{\citenamefont {Willsher}\ \emph {et~al.}(2022)\citenamefont
  {Willsher}, \citenamefont {Liu}, \citenamefont {Moessner},\ and\
  \citenamefont {Knolle}}]{willsher2022measurementinducedphase}%
  \BibitemOpen
  \bibfield  {author} {\bibinfo {author} {\bibfnamefont {J.}~\bibnamefont
  {Willsher}}, \bibinfo {author} {\bibfnamefont {S.-W.}\ \bibnamefont {Liu}},
  \bibinfo {author} {\bibfnamefont {R.}~\bibnamefont {Moessner}},\ and\
  \bibinfo {author} {\bibfnamefont {J.}~\bibnamefont {Knolle}},\ }\href
  {https://doi.org/10.1103/PhysRevB.106.024305} {\bibfield  {journal} {\bibinfo
   {journal} {Phys. Rev. B}\ }\textbf {\bibinfo {volume} {106}},\ \bibinfo
  {pages} {024305} (\bibinfo {year} {2022})}\BibitemShut {NoStop}%
\bibitem [{\citenamefont {Pizzi}\ \emph {et~al.}(2022)\citenamefont {Pizzi},
  \citenamefont {Malz}, \citenamefont {Nunnenkamp},\ and\ \citenamefont
  {Knolle}}]{pizzi2022bridgingthegap}%
  \BibitemOpen
  \bibfield  {author} {\bibinfo {author} {\bibfnamefont {A.}~\bibnamefont
  {Pizzi}}, \bibinfo {author} {\bibfnamefont {D.}~\bibnamefont {Malz}},
  \bibinfo {author} {\bibfnamefont {A.}~\bibnamefont {Nunnenkamp}},\ and\
  \bibinfo {author} {\bibfnamefont {J.}~\bibnamefont {Knolle}},\ }\href
  {https://doi.org/10.1103/PhysRevB.106.214303} {\bibfield  {journal} {\bibinfo
   {journal} {Phys. Rev. B}\ }\textbf {\bibinfo {volume} {106}},\ \bibinfo
  {pages} {214303} (\bibinfo {year} {2022})}\BibitemShut {NoStop}%
\bibitem [{\citenamefont {Lyons}\ \emph {et~al.}()\citenamefont {Lyons},
  \citenamefont {Choi},\ and\ \citenamefont
  {Altman}}]{lyons2022auniversalcrossover}%
  \BibitemOpen
  \bibfield  {author} {\bibinfo {author} {\bibfnamefont {A.}~\bibnamefont
  {Lyons}}, \bibinfo {author} {\bibfnamefont {S.}~\bibnamefont {Choi}},\ and\
  \bibinfo {author} {\bibfnamefont {E.}~\bibnamefont {Altman}},\ }\href@noop {}
  {}\Eprint {https://arxiv.org/abs/2208.02217} {arXiv:2208.02217} \BibitemShut
  {NoStop}%
\bibitem [{\citenamefont {Zhang}\ \emph {et~al.}(2020)\citenamefont {Zhang},
  \citenamefont {Reyes}, \citenamefont {Kourtis}, \citenamefont {Chamon},
  \citenamefont {Mucciolo},\ and\ \citenamefont
  {Ruckenstein}}]{zhang2020nonuniversalentanglementlevel}%
  \BibitemOpen
  \bibfield  {author} {\bibinfo {author} {\bibfnamefont {L.}~\bibnamefont
  {Zhang}}, \bibinfo {author} {\bibfnamefont {J.~A.}\ \bibnamefont {Reyes}},
  \bibinfo {author} {\bibfnamefont {S.}~\bibnamefont {Kourtis}}, \bibinfo
  {author} {\bibfnamefont {C.}~\bibnamefont {Chamon}}, \bibinfo {author}
  {\bibfnamefont {E.~R.}\ \bibnamefont {Mucciolo}},\ and\ \bibinfo {author}
  {\bibfnamefont {A.~E.}\ \bibnamefont {Ruckenstein}},\ }\href
  {https://doi.org/10.1103/PhysRevB.101.235104} {\bibfield  {journal} {\bibinfo
   {journal} {Phys. Rev. B}\ }\textbf {\bibinfo {volume} {101}},\ \bibinfo
  {pages} {235104} (\bibinfo {year} {2020})}\BibitemShut {NoStop}%
\bibitem [{\citenamefont {Zhang}\ \emph {et~al.}(2021)\citenamefont {Zhang},
  \citenamefont {Jian}, \citenamefont {Liu},\ and\ \citenamefont
  {Chen}}]{zhang2021emergentreplica}%
  \BibitemOpen
  \bibfield  {author} {\bibinfo {author} {\bibfnamefont {P.}~\bibnamefont
  {Zhang}}, \bibinfo {author} {\bibfnamefont {S.-K.}\ \bibnamefont {Jian}},
  \bibinfo {author} {\bibfnamefont {C.}~\bibnamefont {Liu}},\ and\ \bibinfo
  {author} {\bibfnamefont {X.}~\bibnamefont {Chen}},\ }\href
  {https://doi.org/10.22331/q-2021-11-16-579} {\bibfield  {journal} {\bibinfo
  {journal} {{Quantum}}\ }\textbf {\bibinfo {volume} {5}},\ \bibinfo {pages}
  {579} (\bibinfo {year} {2021})}\BibitemShut {NoStop}%
\bibitem [{\citenamefont {Zhang}\ \emph {et~al.}(2022)\citenamefont {Zhang},
  \citenamefont {Liu}, \citenamefont {Jian},\ and\ \citenamefont
  {Chen}}]{zhang2022universalentanglementtransitions}%
  \BibitemOpen
  \bibfield  {author} {\bibinfo {author} {\bibfnamefont {P.}~\bibnamefont
  {Zhang}}, \bibinfo {author} {\bibfnamefont {C.}~\bibnamefont {Liu}}, \bibinfo
  {author} {\bibfnamefont {S.-K.}\ \bibnamefont {Jian}},\ and\ \bibinfo
  {author} {\bibfnamefont {X.}~\bibnamefont {Chen}},\ }\href
  {https://doi.org/10.22331/q-2022-05-27-723} {\bibfield  {journal} {\bibinfo
  {journal} {{Quantum}}\ }\textbf {\bibinfo {volume} {6}},\ \bibinfo {pages}
  {723} (\bibinfo {year} {2022})}\BibitemShut {NoStop}%
\bibitem [{\citenamefont {Zhou}\ and\ \citenamefont
  {Chen}(2021)}]{zhou2021nonunitaryentanglementdynamics}%
  \BibitemOpen
  \bibfield  {author} {\bibinfo {author} {\bibfnamefont {T.}~\bibnamefont
  {Zhou}}\ and\ \bibinfo {author} {\bibfnamefont {X.}~\bibnamefont {Chen}},\
  }\href {https://doi.org/10.1103/PhysRevB.104.L180301} {\bibfield  {journal}
  {\bibinfo  {journal} {Phys. Rev. B}\ }\textbf {\bibinfo {volume} {104}},\
  \bibinfo {pages} {L180301} (\bibinfo {year} {2021})}\BibitemShut {NoStop}%
\bibitem [{\citenamefont {Bentsen}\ \emph {et~al.}(2021)\citenamefont
  {Bentsen}, \citenamefont {Sahu},\ and\ \citenamefont
  {Swingle}}]{bentsen2021measurementinducedpurification}%
  \BibitemOpen
  \bibfield  {author} {\bibinfo {author} {\bibfnamefont {G.~S.}\ \bibnamefont
  {Bentsen}}, \bibinfo {author} {\bibfnamefont {S.}~\bibnamefont {Sahu}},\ and\
  \bibinfo {author} {\bibfnamefont {B.}~\bibnamefont {Swingle}},\ }\href
  {https://doi.org/10.1103/PhysRevB.104.094304} {\bibfield  {journal} {\bibinfo
   {journal} {Phys. Rev. B}\ }\textbf {\bibinfo {volume} {104}},\ \bibinfo
  {pages} {094304} (\bibinfo {year} {2021})}\BibitemShut {NoStop}%
\bibitem [{\citenamefont {Yang}\ \emph {et~al.}(2022)\citenamefont {Yang},
  \citenamefont {Li}, \citenamefont {Fisher},\ and\ \citenamefont
  {Chen}}]{yang2022entanglementphasetransitions}%
  \BibitemOpen
  \bibfield  {author} {\bibinfo {author} {\bibfnamefont {Z.-C.}\ \bibnamefont
  {Yang}}, \bibinfo {author} {\bibfnamefont {Y.}~\bibnamefont {Li}}, \bibinfo
  {author} {\bibfnamefont {M.~P.~A.}\ \bibnamefont {Fisher}},\ and\ \bibinfo
  {author} {\bibfnamefont {X.}~\bibnamefont {Chen}},\ }\href
  {https://doi.org/10.1103/PhysRevB.105.104306} {\bibfield  {journal} {\bibinfo
   {journal} {Phys. Rev. B}\ }\textbf {\bibinfo {volume} {105}},\ \bibinfo
  {pages} {104306} (\bibinfo {year} {2022})}\BibitemShut {NoStop}%
\bibitem [{\citenamefont {Rossini}\ and\ \citenamefont
  {Vicari}(2020)}]{rossini2020measurementinduceddynamics}%
  \BibitemOpen
  \bibfield  {author} {\bibinfo {author} {\bibfnamefont {D.}~\bibnamefont
  {Rossini}}\ and\ \bibinfo {author} {\bibfnamefont {E.}~\bibnamefont
  {Vicari}},\ }\href {https://doi.org/10.1103/PhysRevB.102.035119} {\bibfield
  {journal} {\bibinfo  {journal} {Phys. Rev. B}\ }\textbf {\bibinfo {volume}
  {102}},\ \bibinfo {pages} {035119} (\bibinfo {year} {2020})}\BibitemShut
  {NoStop}%
\bibitem [{\citenamefont {Medina}\ \emph {et~al.}(2021)\citenamefont {Medina},
  \citenamefont {Vasseur},\ and\ \citenamefont
  {Serbyn}}]{medina2021entanglementtransitionsfrom}%
  \BibitemOpen
  \bibfield  {author} {\bibinfo {author} {\bibfnamefont {R.}~\bibnamefont
  {Medina}}, \bibinfo {author} {\bibfnamefont {R.}~\bibnamefont {Vasseur}},\
  and\ \bibinfo {author} {\bibfnamefont {M.}~\bibnamefont {Serbyn}},\ }\href
  {https://doi.org/10.1103/PhysRevB.104.104205} {\bibfield  {journal} {\bibinfo
   {journal} {Phys. Rev. B}\ }\textbf {\bibinfo {volume} {104}},\ \bibinfo
  {pages} {104205} (\bibinfo {year} {2021})}\BibitemShut {NoStop}%
\bibitem [{\citenamefont {Lunt}\ and\ \citenamefont
  {Pal}(2020)}]{lunt2020measurementinducedentanglement}%
  \BibitemOpen
  \bibfield  {author} {\bibinfo {author} {\bibfnamefont {O.}~\bibnamefont
  {Lunt}}\ and\ \bibinfo {author} {\bibfnamefont {A.}~\bibnamefont {Pal}},\
  }\href {https://doi.org/10.1103/PhysRevResearch.2.043072} {\bibfield
  {journal} {\bibinfo  {journal} {Phys. Rev. Res.}\ }\textbf {\bibinfo {volume}
  {2}},\ \bibinfo {pages} {043072} (\bibinfo {year} {2020})}\BibitemShut
  {NoStop}%
\bibitem [{\citenamefont {Kelly}\ \emph {et~al.}()\citenamefont {Kelly},
  \citenamefont {Poschinger}, \citenamefont {Schmidt-Kaler}, \citenamefont
  {Fisher},\ and\ \citenamefont {Marino}}]{kelly2022coherencerequirementsfor}%
  \BibitemOpen
  \bibfield  {author} {\bibinfo {author} {\bibfnamefont {S.~P.}\ \bibnamefont
  {Kelly}}, \bibinfo {author} {\bibfnamefont {U.}~\bibnamefont {Poschinger}},
  \bibinfo {author} {\bibfnamefont {F.}~\bibnamefont {Schmidt-Kaler}}, \bibinfo
  {author} {\bibfnamefont {M.~P.~A.}\ \bibnamefont {Fisher}},\ and\ \bibinfo
  {author} {\bibfnamefont {J.}~\bibnamefont {Marino}},\ }\href@noop {}
  {}\Eprint {https://arxiv.org/abs/2210.11547} {arXiv:2210.11547} \BibitemShut
  {NoStop}%
\bibitem [{\citenamefont {Szyniszewski}\ \emph {et~al.}(2019)\citenamefont
  {Szyniszewski}, \citenamefont {Romito},\ and\ \citenamefont
  {Schomerus}}]{szyniszewski2019entanglementtransitionfrom}%
  \BibitemOpen
  \bibfield  {author} {\bibinfo {author} {\bibfnamefont {M.}~\bibnamefont
  {Szyniszewski}}, \bibinfo {author} {\bibfnamefont {A.}~\bibnamefont
  {Romito}},\ and\ \bibinfo {author} {\bibfnamefont {H.}~\bibnamefont
  {Schomerus}},\ }\href {https://doi.org/10.1103/PhysRevB.100.064204}
  {\bibfield  {journal} {\bibinfo  {journal} {Phys. Rev. B}\ }\textbf {\bibinfo
  {volume} {100}},\ \bibinfo {pages} {064204} (\bibinfo {year}
  {2019})}\BibitemShut {NoStop}%
\bibitem [{\citenamefont {Szyniszewski}\ \emph {et~al.}(2020)\citenamefont
  {Szyniszewski}, \citenamefont {Romito},\ and\ \citenamefont
  {Schomerus}}]{szyniszewski2020universalityofentanglement}%
  \BibitemOpen
  \bibfield  {author} {\bibinfo {author} {\bibfnamefont {M.}~\bibnamefont
  {Szyniszewski}}, \bibinfo {author} {\bibfnamefont {A.}~\bibnamefont
  {Romito}},\ and\ \bibinfo {author} {\bibfnamefont {H.}~\bibnamefont
  {Schomerus}},\ }\href {https://doi.org/10.1103/PhysRevLett.125.210602}
  {\bibfield  {journal} {\bibinfo  {journal} {Phys. Rev. Lett.}\ }\textbf
  {\bibinfo {volume} {125}},\ \bibinfo {pages} {210602} (\bibinfo {year}
  {2020})}\BibitemShut {NoStop}%
\bibitem [{\citenamefont {Tang}\ and\ \citenamefont
  {Zhu}(2020)}]{tang2020measurementinducedphase}%
  \BibitemOpen
  \bibfield  {author} {\bibinfo {author} {\bibfnamefont {Q.}~\bibnamefont
  {Tang}}\ and\ \bibinfo {author} {\bibfnamefont {W.}~\bibnamefont {Zhu}},\
  }\href {https://doi.org/10.1103/PhysRevResearch.2.013022} {\bibfield
  {journal} {\bibinfo  {journal} {Phys. Rev. Res.}\ }\textbf {\bibinfo {volume}
  {2}},\ \bibinfo {pages} {013022} (\bibinfo {year} {2020})}\BibitemShut
  {NoStop}%
\bibitem [{\citenamefont {Iadecola}\ \emph {et~al.}()\citenamefont {Iadecola},
  \citenamefont {Ganeshan}, \citenamefont {Pixley},\ and\ \citenamefont
  {Wilson}}]{iadecola2022dynamicalentanglementtransition}%
  \BibitemOpen
  \bibfield  {author} {\bibinfo {author} {\bibfnamefont {T.}~\bibnamefont
  {Iadecola}}, \bibinfo {author} {\bibfnamefont {S.}~\bibnamefont {Ganeshan}},
  \bibinfo {author} {\bibfnamefont {J.~H.}\ \bibnamefont {Pixley}},\ and\
  \bibinfo {author} {\bibfnamefont {J.~H.}\ \bibnamefont {Wilson}},\
  }\href@noop {} {}\Eprint {https://arxiv.org/abs/2207.12415}
  {arXiv:2207.12415} \BibitemShut {NoStop}%
\bibitem [{\citenamefont {O'Dea}\ \emph {et~al.}()\citenamefont {O'Dea},
  \citenamefont {Morningstar}, \citenamefont {Gopalakrishnan},\ and\
  \citenamefont {Khemani}}]{odea2022entanglementandabsorbing}%
  \BibitemOpen
  \bibfield  {author} {\bibinfo {author} {\bibfnamefont {N.}~\bibnamefont
  {O'Dea}}, \bibinfo {author} {\bibfnamefont {A.}~\bibnamefont {Morningstar}},
  \bibinfo {author} {\bibfnamefont {S.}~\bibnamefont {Gopalakrishnan}},\ and\
  \bibinfo {author} {\bibfnamefont {V.}~\bibnamefont {Khemani}},\ }\href@noop
  {} {}\Eprint {https://arxiv.org/abs/2211.12526} {arXiv:2211.12526}
  \BibitemShut {NoStop}%
\bibitem [{\citenamefont {Ravindranath}\ \emph {et~al.}()\citenamefont
  {Ravindranath}, \citenamefont {Han}, \citenamefont {Yang},\ and\
  \citenamefont {Chen}}]{ravindranath2022entanglementsteeringin}%
  \BibitemOpen
  \bibfield  {author} {\bibinfo {author} {\bibfnamefont {V.}~\bibnamefont
  {Ravindranath}}, \bibinfo {author} {\bibfnamefont {Y.}~\bibnamefont {Han}},
  \bibinfo {author} {\bibfnamefont {Z.-C.}\ \bibnamefont {Yang}},\ and\
  \bibinfo {author} {\bibfnamefont {X.}~\bibnamefont {Chen}},\ }\href@noop {}
  {}\Eprint {https://arxiv.org/abs/2211.05162} {arXiv:2211.05162} \BibitemShut
  {NoStop}%
\bibitem [{\citenamefont {Piroli}\ \emph {et~al.}()\citenamefont {Piroli},
  \citenamefont {Li}, \citenamefont {Vasseur},\ and\ \citenamefont
  {Nahum}}]{piroli2022trivialityofquantum}%
  \BibitemOpen
  \bibfield  {author} {\bibinfo {author} {\bibfnamefont {L.}~\bibnamefont
  {Piroli}}, \bibinfo {author} {\bibfnamefont {Y.}~\bibnamefont {Li}}, \bibinfo
  {author} {\bibfnamefont {R.}~\bibnamefont {Vasseur}},\ and\ \bibinfo {author}
  {\bibfnamefont {A.}~\bibnamefont {Nahum}},\ }\href
  {https://doi.org/10.48550/ARXIV.2212.14026} {}\Eprint
  {https://arxiv.org/abs/2212.14026} {arXiv:2212.14026} \BibitemShut {NoStop}%
\bibitem [{\citenamefont {Sierant}\ and\ \citenamefont
  {Turkeshi}()}]{sierant2022controllingentanglementat}%
  \BibitemOpen
  \bibfield  {author} {\bibinfo {author} {\bibfnamefont {P.}~\bibnamefont
  {Sierant}}\ and\ \bibinfo {author} {\bibfnamefont {X.}~\bibnamefont
  {Turkeshi}},\ }\href {https://doi.org/10.48550/ARXIV.2212.13823} {}\Eprint
  {https://arxiv.org/abs/2212.13823} {arXiv:2212.13823} \BibitemShut {NoStop}%
\bibitem [{\citenamefont {Vijay}()}]{vijay2020measurementdrivenphase}%
  \BibitemOpen
  \bibfield  {author} {\bibinfo {author} {\bibfnamefont {S.}~\bibnamefont
  {Vijay}},\ }\href@noop {} {}\Eprint {https://arxiv.org/abs/2005.03052}
  {arXiv:2005.03052} \BibitemShut {NoStop}%
\bibitem [{\citenamefont {Fan}\ \emph {et~al.}(2021)\citenamefont {Fan},
  \citenamefont {Vijay}, \citenamefont {Vishwanath},\ and\ \citenamefont
  {You}}]{fan2021selforganizederror}%
  \BibitemOpen
  \bibfield  {author} {\bibinfo {author} {\bibfnamefont {R.}~\bibnamefont
  {Fan}}, \bibinfo {author} {\bibfnamefont {S.}~\bibnamefont {Vijay}}, \bibinfo
  {author} {\bibfnamefont {A.}~\bibnamefont {Vishwanath}},\ and\ \bibinfo
  {author} {\bibfnamefont {Y.-Z.}\ \bibnamefont {You}},\ }\href
  {https://doi.org/10.1103/PhysRevB.103.174309} {\bibfield  {journal} {\bibinfo
   {journal} {Phys. Rev. B}\ }\textbf {\bibinfo {volume} {103}},\ \bibinfo
  {pages} {174309} (\bibinfo {year} {2021})}\BibitemShut {NoStop}%
\bibitem [{\citenamefont {Li}\ \emph {et~al.}(2021)\citenamefont {Li},
  \citenamefont {Chen}, \citenamefont {Ludwig},\ and\ \citenamefont
  {Fisher}}]{li2021conformal}%
  \BibitemOpen
  \bibfield  {author} {\bibinfo {author} {\bibfnamefont {Y.}~\bibnamefont
  {Li}}, \bibinfo {author} {\bibfnamefont {X.}~\bibnamefont {Chen}}, \bibinfo
  {author} {\bibfnamefont {A.~W.~W.}\ \bibnamefont {Ludwig}},\ and\ \bibinfo
  {author} {\bibfnamefont {M.~P.~A.}\ \bibnamefont {Fisher}},\ }\href
  {https://doi.org/10.1103/PhysRevB.104.104305} {\bibfield  {journal} {\bibinfo
   {journal} {Phys. Rev. B}\ }\textbf {\bibinfo {volume} {104}},\ \bibinfo
  {pages} {104305} (\bibinfo {year} {2021})}\BibitemShut {NoStop}%
\bibitem [{\citenamefont {Ippoliti}\ \emph {et~al.}(2021)\citenamefont
  {Ippoliti}, \citenamefont {Gullans}, \citenamefont {Gopalakrishnan},
  \citenamefont {Huse},\ and\ \citenamefont
  {Khemani}}]{ippoliti2021entanglementphasetransitions}%
  \BibitemOpen
  \bibfield  {author} {\bibinfo {author} {\bibfnamefont {M.}~\bibnamefont
  {Ippoliti}}, \bibinfo {author} {\bibfnamefont {M.~J.}\ \bibnamefont
  {Gullans}}, \bibinfo {author} {\bibfnamefont {S.}~\bibnamefont
  {Gopalakrishnan}}, \bibinfo {author} {\bibfnamefont {D.~A.}\ \bibnamefont
  {Huse}},\ and\ \bibinfo {author} {\bibfnamefont {V.}~\bibnamefont
  {Khemani}},\ }\href {https://doi.org/10.1103/PhysRevX.11.011030} {\bibfield
  {journal} {\bibinfo  {journal} {Phys. Rev. X}\ }\textbf {\bibinfo {volume}
  {11}},\ \bibinfo {pages} {011030} (\bibinfo {year} {2021})}\BibitemShut
  {NoStop}%
\bibitem [{\citenamefont {Ippoliti}\ \emph {et~al.}(2022)\citenamefont
  {Ippoliti}, \citenamefont {Rakovszky},\ and\ \citenamefont
  {Khemani}}]{ippoliti2022fractallogarithmicand}%
  \BibitemOpen
  \bibfield  {author} {\bibinfo {author} {\bibfnamefont {M.}~\bibnamefont
  {Ippoliti}}, \bibinfo {author} {\bibfnamefont {T.}~\bibnamefont
  {Rakovszky}},\ and\ \bibinfo {author} {\bibfnamefont {V.}~\bibnamefont
  {Khemani}},\ }\href {https://doi.org/10.1103/PhysRevX.12.011045} {\bibfield
  {journal} {\bibinfo  {journal} {Phys. Rev. X}\ }\textbf {\bibinfo {volume}
  {12}},\ \bibinfo {pages} {011045} (\bibinfo {year} {2022})}\BibitemShut
  {NoStop}%
\bibitem [{\citenamefont {Klocke}\ and\ \citenamefont
  {Buchhold}(2022)}]{klocke2022topologicalorderand}%
  \BibitemOpen
  \bibfield  {author} {\bibinfo {author} {\bibfnamefont {K.}~\bibnamefont
  {Klocke}}\ and\ \bibinfo {author} {\bibfnamefont {M.}~\bibnamefont
  {Buchhold}},\ }\href {https://doi.org/10.1103/PhysRevB.106.104307} {\bibfield
   {journal} {\bibinfo  {journal} {Phys. Rev. B}\ }\textbf {\bibinfo {volume}
  {106}},\ \bibinfo {pages} {104307} (\bibinfo {year} {2022})}\BibitemShut
  {NoStop}%
\bibitem [{\citenamefont {Lu}\ and\ \citenamefont
  {Grover}(2021)}]{lu2021spacetimeduality}%
  \BibitemOpen
  \bibfield  {author} {\bibinfo {author} {\bibfnamefont {T.-C.}\ \bibnamefont
  {Lu}}\ and\ \bibinfo {author} {\bibfnamefont {T.}~\bibnamefont {Grover}},\
  }\href {https://doi.org/10.1103/PRXQuantum.2.040319} {\bibfield  {journal}
  {\bibinfo  {journal} {PRX Quantum}\ }\textbf {\bibinfo {volume} {2}},\
  \bibinfo {pages} {040319} (\bibinfo {year} {2021})}\BibitemShut {NoStop}%
\bibitem [{\citenamefont {Ippoliti}\ and\ \citenamefont
  {Khemani}(2021)}]{ippoliti2021postselectionfreeentanglement}%
  \BibitemOpen
  \bibfield  {author} {\bibinfo {author} {\bibfnamefont {M.}~\bibnamefont
  {Ippoliti}}\ and\ \bibinfo {author} {\bibfnamefont {V.}~\bibnamefont
  {Khemani}},\ }\href {https://doi.org/10.1103/PhysRevLett.126.060501}
  {\bibfield  {journal} {\bibinfo  {journal} {Phys. Rev. Lett.}\ }\textbf
  {\bibinfo {volume} {126}},\ \bibinfo {pages} {060501} (\bibinfo {year}
  {2021})}\BibitemShut {NoStop}%
\bibitem [{\citenamefont {Li}\ and\ \citenamefont
  {Fisher}()}]{li2021robustdecodingin}%
  \BibitemOpen
  \bibfield  {author} {\bibinfo {author} {\bibfnamefont {Y.}~\bibnamefont
  {Li}}\ and\ \bibinfo {author} {\bibfnamefont {M.~P.~A.}\ \bibnamefont
  {Fisher}},\ }\href@noop {} {}\Eprint {https://arxiv.org/abs/2108.04274}
  {arXiv:2108.04274} \BibitemShut {NoStop}%
\bibitem [{\citenamefont {Li}\ \emph {et~al.}()\citenamefont {Li},
  \citenamefont {Vasseur}, \citenamefont {Fisher},\ and\ \citenamefont
  {Ludwig}}]{li2021statisticalmechanicsmodel}%
  \BibitemOpen
  \bibfield  {author} {\bibinfo {author} {\bibfnamefont {Y.}~\bibnamefont
  {Li}}, \bibinfo {author} {\bibfnamefont {R.}~\bibnamefont {Vasseur}},
  \bibinfo {author} {\bibfnamefont {M.~P.~A.}\ \bibnamefont {Fisher}},\ and\
  \bibinfo {author} {\bibfnamefont {A.~W.~W.}\ \bibnamefont {Ludwig}},\
  }\href@noop {} {}\Eprint {https://arxiv.org/abs/2110.02988}
  {arXiv:2110.02988} \BibitemShut {NoStop}%
\bibitem [{\citenamefont {{Y. Li, S. Vijay, and M. P. A.
  Fisher}}()}]{li2021entanglementdomainwalls}%
  \BibitemOpen
  \bibfield  {author} {\bibinfo {author} {\bibnamefont {{Y. Li, S. Vijay, and
  M. P. A. Fisher}}},\ }\href@noop {} {}\Eprint
  {https://arxiv.org/abs/2105.13352} {arXiv:2105.13352} \BibitemShut {NoStop}%
\bibitem [{\citenamefont {Li}\ and\ \citenamefont
  {Fisher}(2021)}]{li2021statisticalmechanicsof}%
  \BibitemOpen
  \bibfield  {author} {\bibinfo {author} {\bibfnamefont {Y.}~\bibnamefont
  {Li}}\ and\ \bibinfo {author} {\bibfnamefont {M.~P.~A.}\ \bibnamefont
  {Fisher}},\ }\href {https://doi.org/10.1103/PhysRevB.103.104306} {\bibfield
  {journal} {\bibinfo  {journal} {Phys. Rev. B}\ }\textbf {\bibinfo {volume}
  {103}},\ \bibinfo {pages} {104306} (\bibinfo {year} {2021})}\BibitemShut
  {NoStop}%
\bibitem [{\citenamefont {Feng}\ \emph {et~al.}()\citenamefont {Feng},
  \citenamefont {Skinner},\ and\ \citenamefont
  {Nahum}}]{feng2022measurementinducedphase}%
  \BibitemOpen
  \bibfield  {author} {\bibinfo {author} {\bibfnamefont {X.}~\bibnamefont
  {Feng}}, \bibinfo {author} {\bibfnamefont {B.}~\bibnamefont {Skinner}},\ and\
  \bibinfo {author} {\bibfnamefont {A.}~\bibnamefont {Nahum}},\ }\href@noop {}
  {}\Eprint {https://arxiv.org/abs/2210.07264} {arXiv:2210.07264} \BibitemShut
  {NoStop}%
\bibitem [{\citenamefont {Barratt}\ \emph {et~al.}(2022)\citenamefont
  {Barratt}, \citenamefont {Agrawal}, \citenamefont {Potter}, \citenamefont
  {Gopalakrishnan},\ and\ \citenamefont {Vasseur}}]{barratt2022transitions}%
  \BibitemOpen
  \bibfield  {author} {\bibinfo {author} {\bibfnamefont {F.}~\bibnamefont
  {Barratt}}, \bibinfo {author} {\bibfnamefont {U.}~\bibnamefont {Agrawal}},
  \bibinfo {author} {\bibfnamefont {A.~C.}\ \bibnamefont {Potter}}, \bibinfo
  {author} {\bibfnamefont {S.}~\bibnamefont {Gopalakrishnan}},\ and\ \bibinfo
  {author} {\bibfnamefont {R.}~\bibnamefont {Vasseur}},\ }\href
  {https://doi.org/10.1103/PhysRevLett.129.200602} {\bibfield  {journal}
  {\bibinfo  {journal} {Phys. Rev. Lett.}\ }\textbf {\bibinfo {volume} {129}},\
  \bibinfo {pages} {200602} (\bibinfo {year} {2022})}\BibitemShut {NoStop}%
\bibitem [{\citenamefont {Zabalo}\ \emph {et~al.}(2022)\citenamefont {Zabalo},
  \citenamefont {Gullans}, \citenamefont {Wilson}, \citenamefont {Vasseur},
  \citenamefont {Ludwig}, \citenamefont {Gopalakrishnan}, \citenamefont
  {Huse},\ and\ \citenamefont {Pixley}}]{zabalo2022operatorscalingdimensions}%
  \BibitemOpen
  \bibfield  {author} {\bibinfo {author} {\bibfnamefont {A.}~\bibnamefont
  {Zabalo}}, \bibinfo {author} {\bibfnamefont {M.~J.}\ \bibnamefont {Gullans}},
  \bibinfo {author} {\bibfnamefont {J.~H.}\ \bibnamefont {Wilson}}, \bibinfo
  {author} {\bibfnamefont {R.}~\bibnamefont {Vasseur}}, \bibinfo {author}
  {\bibfnamefont {A.~W.~W.}\ \bibnamefont {Ludwig}}, \bibinfo {author}
  {\bibfnamefont {S.}~\bibnamefont {Gopalakrishnan}}, \bibinfo {author}
  {\bibfnamefont {D.~A.}\ \bibnamefont {Huse}},\ and\ \bibinfo {author}
  {\bibfnamefont {J.~H.}\ \bibnamefont {Pixley}},\ }\href
  {https://doi.org/10.1103/PhysRevLett.128.050602} {\bibfield  {journal}
  {\bibinfo  {journal} {Phys. Rev. Lett.}\ }\textbf {\bibinfo {volume} {128}},\
  \bibinfo {pages} {050602} (\bibinfo {year} {2022})}\BibitemShut {NoStop}%
\bibitem [{\citenamefont {Sierant}\ and\ \citenamefont
  {Turkeshi}(2022)}]{sierant2022universalbehaviorbeyond}%
  \BibitemOpen
  \bibfield  {author} {\bibinfo {author} {\bibfnamefont {P.}~\bibnamefont
  {Sierant}}\ and\ \bibinfo {author} {\bibfnamefont {X.}~\bibnamefont
  {Turkeshi}},\ }\href {https://doi.org/10.1103/PhysRevLett.128.130605}
  {\bibfield  {journal} {\bibinfo  {journal} {Phys. Rev. Lett.}\ }\textbf
  {\bibinfo {volume} {128}},\ \bibinfo {pages} {130605} (\bibinfo {year}
  {2022})}\BibitemShut {NoStop}%
\bibitem [{\citenamefont {Iaconis}\ \emph {et~al.}(2020)\citenamefont
  {Iaconis}, \citenamefont {Lucas},\ and\ \citenamefont
  {Chen}}]{iaconis2020measurementinducedphase}%
  \BibitemOpen
  \bibfield  {author} {\bibinfo {author} {\bibfnamefont {J.}~\bibnamefont
  {Iaconis}}, \bibinfo {author} {\bibfnamefont {A.}~\bibnamefont {Lucas}},\
  and\ \bibinfo {author} {\bibfnamefont {X.}~\bibnamefont {Chen}},\ }\href
  {https://doi.org/10.1103/PhysRevB.102.224311} {\bibfield  {journal} {\bibinfo
   {journal} {Phys. Rev. B}\ }\textbf {\bibinfo {volume} {102}},\ \bibinfo
  {pages} {224311} (\bibinfo {year} {2020})}\BibitemShut {NoStop}%
\bibitem [{\citenamefont {Han}\ and\ \citenamefont
  {Chen}(2022)}]{han2022measurementinducedcriticality}%
  \BibitemOpen
  \bibfield  {author} {\bibinfo {author} {\bibfnamefont {Y.}~\bibnamefont
  {Han}}\ and\ \bibinfo {author} {\bibfnamefont {X.}~\bibnamefont {Chen}},\
  }\href {https://doi.org/10.1103/PhysRevB.105.064306} {\bibfield  {journal}
  {\bibinfo  {journal} {Phys. Rev. B}\ }\textbf {\bibinfo {volume} {105}},\
  \bibinfo {pages} {064306} (\bibinfo {year} {2022})}\BibitemShut {NoStop}%
\bibitem [{\citenamefont {Liu}\ \emph {et~al.}(2022)\citenamefont {Liu},
  \citenamefont {Zhou},\ and\ \citenamefont
  {Chen}}]{liu2022measurementinducedentanglement}%
  \BibitemOpen
  \bibfield  {author} {\bibinfo {author} {\bibfnamefont {H.}~\bibnamefont
  {Liu}}, \bibinfo {author} {\bibfnamefont {T.}~\bibnamefont {Zhou}},\ and\
  \bibinfo {author} {\bibfnamefont {X.}~\bibnamefont {Chen}},\ }\href
  {https://doi.org/10.1103/PhysRevB.106.144311} {\bibfield  {journal} {\bibinfo
   {journal} {Phys. Rev. B}\ }\textbf {\bibinfo {volume} {106}},\ \bibinfo
  {pages} {144311} (\bibinfo {year} {2022})}\BibitemShut {NoStop}%
\bibitem [{\citenamefont {Sang}\ \emph {et~al.}(2021)\citenamefont {Sang},
  \citenamefont {Li}, \citenamefont {Zhou}, \citenamefont {Chen}, \citenamefont
  {Hsieh},\ and\ \citenamefont {Fisher}}]{sang2021entanglementnegativityat}%
  \BibitemOpen
  \bibfield  {author} {\bibinfo {author} {\bibfnamefont {S.}~\bibnamefont
  {Sang}}, \bibinfo {author} {\bibfnamefont {Y.}~\bibnamefont {Li}}, \bibinfo
  {author} {\bibfnamefont {T.}~\bibnamefont {Zhou}}, \bibinfo {author}
  {\bibfnamefont {X.}~\bibnamefont {Chen}}, \bibinfo {author} {\bibfnamefont
  {T.~H.}\ \bibnamefont {Hsieh}},\ and\ \bibinfo {author} {\bibfnamefont
  {M.~P.}\ \bibnamefont {Fisher}},\ }\href
  {https://doi.org/10.1103/PRXQuantum.2.030313} {\bibfield  {journal} {\bibinfo
   {journal} {PRX Quantum}\ }\textbf {\bibinfo {volume} {2}},\ \bibinfo {pages}
  {030313} (\bibinfo {year} {2021})}\BibitemShut {NoStop}%
\bibitem [{\citenamefont {Shi}\ \emph {et~al.}()\citenamefont {Shi},
  \citenamefont {Dai},\ and\ \citenamefont
  {Lu}}]{shi2020entanglementnegativityat}%
  \BibitemOpen
  \bibfield  {author} {\bibinfo {author} {\bibfnamefont {B.}~\bibnamefont
  {Shi}}, \bibinfo {author} {\bibfnamefont {X.}~\bibnamefont {Dai}},\ and\
  \bibinfo {author} {\bibfnamefont {Y.-M.}\ \bibnamefont {Lu}},\ }\href@noop {}
  {}\Eprint {https://arxiv.org/abs/2012.00040} {arXiv:2012.00040} \BibitemShut
  {NoStop}%
\bibitem [{\citenamefont {Weinstein}\ \emph {et~al.}(2022)\citenamefont
  {Weinstein}, \citenamefont {Bao},\ and\ \citenamefont
  {Altman}}]{weinstein2022measurementinducedpower}%
  \BibitemOpen
  \bibfield  {author} {\bibinfo {author} {\bibfnamefont {Z.}~\bibnamefont
  {Weinstein}}, \bibinfo {author} {\bibfnamefont {Y.}~\bibnamefont {Bao}},\
  and\ \bibinfo {author} {\bibfnamefont {E.}~\bibnamefont {Altman}},\ }\href
  {https://doi.org/10.1103/PhysRevLett.129.080501} {\bibfield  {journal}
  {\bibinfo  {journal} {Phys. Rev. Lett.}\ }\textbf {\bibinfo {volume} {129}},\
  \bibinfo {pages} {080501} (\bibinfo {year} {2022})}\BibitemShut {NoStop}%
\bibitem [{\citenamefont {Turkeshi}\ \emph {et~al.}(2020)\citenamefont
  {Turkeshi}, \citenamefont {Fazio},\ and\ \citenamefont
  {Dalmonte}}]{turkeshi2020measurementinducedcriticality}%
  \BibitemOpen
  \bibfield  {author} {\bibinfo {author} {\bibfnamefont {X.}~\bibnamefont
  {Turkeshi}}, \bibinfo {author} {\bibfnamefont {R.}~\bibnamefont {Fazio}},\
  and\ \bibinfo {author} {\bibfnamefont {M.}~\bibnamefont {Dalmonte}},\ }\href
  {https://doi.org/10.1103/PhysRevB.102.014315} {\bibfield  {journal} {\bibinfo
   {journal} {Phys. Rev. B}\ }\textbf {\bibinfo {volume} {102}},\ \bibinfo
  {pages} {014315} (\bibinfo {year} {2020})}\BibitemShut {NoStop}%
\bibitem [{\citenamefont
  {Turkeshi}(2022)}]{turkeshi2022measurementinducedcriticality}%
  \BibitemOpen
  \bibfield  {author} {\bibinfo {author} {\bibfnamefont {X.}~\bibnamefont
  {Turkeshi}},\ }\href {https://doi.org/10.1103/PhysRevB.106.144313} {\bibfield
   {journal} {\bibinfo  {journal} {Phys. Rev. B}\ }\textbf {\bibinfo {volume}
  {106}},\ \bibinfo {pages} {144313} (\bibinfo {year} {2022})}\BibitemShut
  {NoStop}%
\bibitem [{\citenamefont {Weinstein}\ \emph {et~al.}()\citenamefont
  {Weinstein}, \citenamefont {Kelly}, \citenamefont {Marino},\ and\
  \citenamefont {Altman}}]{weinstein2022scramblngtransitionin}%
  \BibitemOpen
  \bibfield  {author} {\bibinfo {author} {\bibfnamefont {Z.}~\bibnamefont
  {Weinstein}}, \bibinfo {author} {\bibfnamefont {S.~P.}\ \bibnamefont
  {Kelly}}, \bibinfo {author} {\bibfnamefont {J.}~\bibnamefont {Marino}},\ and\
  \bibinfo {author} {\bibfnamefont {E.}~\bibnamefont {Altman}},\ }\href@noop {}
  {}\Eprint {https://arxiv.org/abs/2210.14242} {arXiv:2210.14242} \BibitemShut
  {NoStop}%
\bibitem [{\citenamefont {Zabalo}\ \emph {et~al.}()\citenamefont {Zabalo},
  \citenamefont {Wilson}, \citenamefont {Gullans}, \citenamefont {Vasseur},
  \citenamefont {Gopalakrishnan}, \citenamefont {Huse},\ and\ \citenamefont
  {Pixley}}]{zabalo2022infiniterandomnesscriticality}%
  \BibitemOpen
  \bibfield  {author} {\bibinfo {author} {\bibfnamefont {A.}~\bibnamefont
  {Zabalo}}, \bibinfo {author} {\bibfnamefont {J.~H.}\ \bibnamefont {Wilson}},
  \bibinfo {author} {\bibfnamefont {M.~J.}\ \bibnamefont {Gullans}}, \bibinfo
  {author} {\bibfnamefont {R.}~\bibnamefont {Vasseur}}, \bibinfo {author}
  {\bibfnamefont {S.}~\bibnamefont {Gopalakrishnan}}, \bibinfo {author}
  {\bibfnamefont {D.~A.}\ \bibnamefont {Huse}},\ and\ \bibinfo {author}
  {\bibfnamefont {J.~H.}\ \bibnamefont {Pixley}},\ }\href@noop {} {}\Eprint
  {https://arxiv.org/abs/2205.14002} {arXiv:2205.14002} \BibitemShut {NoStop}%
\bibitem [{\citenamefont {Jian}\ \emph
  {et~al.}(2021{\natexlab{b}})\citenamefont {Jian}, \citenamefont {Liu},
  \citenamefont {Chen}, \citenamefont {Swingle},\ and\ \citenamefont
  {Zhang}}]{jian2021measurementinducedphase}%
  \BibitemOpen
  \bibfield  {author} {\bibinfo {author} {\bibfnamefont {S.-K.}\ \bibnamefont
  {Jian}}, \bibinfo {author} {\bibfnamefont {C.}~\bibnamefont {Liu}}, \bibinfo
  {author} {\bibfnamefont {X.}~\bibnamefont {Chen}}, \bibinfo {author}
  {\bibfnamefont {B.}~\bibnamefont {Swingle}},\ and\ \bibinfo {author}
  {\bibfnamefont {P.}~\bibnamefont {Zhang}},\ }\href
  {https://doi.org/10.1103/PhysRevLett.127.140601} {\bibfield  {journal}
  {\bibinfo  {journal} {Phys. Rev. Lett.}\ }\textbf {\bibinfo {volume} {127}},\
  \bibinfo {pages} {140601} (\bibinfo {year} {2021}{\natexlab{b}})}\BibitemShut
  {NoStop}%
\bibitem [{\citenamefont {Lopez-Piqueres}\ \emph {et~al.}(2020)\citenamefont
  {Lopez-Piqueres}, \citenamefont {Ware},\ and\ \citenamefont
  {Vasseur}}]{lopezpiqueres2020meanfieldentanglement}%
  \BibitemOpen
  \bibfield  {author} {\bibinfo {author} {\bibfnamefont {J.}~\bibnamefont
  {Lopez-Piqueres}}, \bibinfo {author} {\bibfnamefont {B.}~\bibnamefont
  {Ware}},\ and\ \bibinfo {author} {\bibfnamefont {R.}~\bibnamefont
  {Vasseur}},\ }\href {https://doi.org/10.1103/PhysRevB.102.064202} {\bibfield
  {journal} {\bibinfo  {journal} {Phys. Rev. B}\ }\textbf {\bibinfo {volume}
  {102}},\ \bibinfo {pages} {064202} (\bibinfo {year} {2020})}\BibitemShut
  {NoStop}%
\bibitem [{\citenamefont {Vasseur}\ \emph {et~al.}(2019)\citenamefont
  {Vasseur}, \citenamefont {Potter}, \citenamefont {You},\ and\ \citenamefont
  {Ludwig}}]{vasseur2019entanglementtransitionsfrom}%
  \BibitemOpen
  \bibfield  {author} {\bibinfo {author} {\bibfnamefont {R.}~\bibnamefont
  {Vasseur}}, \bibinfo {author} {\bibfnamefont {A.~C.}\ \bibnamefont {Potter}},
  \bibinfo {author} {\bibfnamefont {Y.-Z.}\ \bibnamefont {You}},\ and\ \bibinfo
  {author} {\bibfnamefont {A.~W.~W.}\ \bibnamefont {Ludwig}},\ }\href
  {https://doi.org/10.1103/PhysRevB.100.134203} {\bibfield  {journal} {\bibinfo
   {journal} {Phys. Rev. B}\ }\textbf {\bibinfo {volume} {100}},\ \bibinfo
  {pages} {134203} (\bibinfo {year} {2019})}\BibitemShut {NoStop}%
\bibitem [{\citenamefont {Jian}\ \emph {et~al.}(2020)\citenamefont {Jian},
  \citenamefont {You}, \citenamefont {Vasseur},\ and\ \citenamefont
  {Ludwig}}]{jian2020measurementinducedcriticality}%
  \BibitemOpen
  \bibfield  {author} {\bibinfo {author} {\bibfnamefont {C.-M.}\ \bibnamefont
  {Jian}}, \bibinfo {author} {\bibfnamefont {Y.-Z.}\ \bibnamefont {You}},
  \bibinfo {author} {\bibfnamefont {R.}~\bibnamefont {Vasseur}},\ and\ \bibinfo
  {author} {\bibfnamefont {A.~W.~W.}\ \bibnamefont {Ludwig}},\ }\href
  {https://doi.org/10.1103/PhysRevB.101.104302} {\bibfield  {journal} {\bibinfo
   {journal} {Phys. Rev. B}\ }\textbf {\bibinfo {volume} {101}},\ \bibinfo
  {pages} {104302} (\bibinfo {year} {2020})}\BibitemShut {NoStop}%
\bibitem [{\citenamefont {Nahum}\ \emph {et~al.}(2021)\citenamefont {Nahum},
  \citenamefont {Roy}, \citenamefont {Skinner},\ and\ \citenamefont
  {Ruhman}}]{nahum2921measurementandentanglement}%
  \BibitemOpen
  \bibfield  {author} {\bibinfo {author} {\bibfnamefont {A.}~\bibnamefont
  {Nahum}}, \bibinfo {author} {\bibfnamefont {S.}~\bibnamefont {Roy}}, \bibinfo
  {author} {\bibfnamefont {B.}~\bibnamefont {Skinner}},\ and\ \bibinfo {author}
  {\bibfnamefont {J.}~\bibnamefont {Ruhman}},\ }\href
  {https://doi.org/10.1103/PRXQuantum.2.010352} {\bibfield  {journal} {\bibinfo
   {journal} {PRX Quantum}\ }\textbf {\bibinfo {volume} {2}},\ \bibinfo {pages}
  {010352} (\bibinfo {year} {2021})}\BibitemShut {NoStop}%
\bibitem [{\citenamefont {Nahum}\ and\ \citenamefont
  {Wiese}()}]{nahum2023renormalizationgroupfor}%
  \BibitemOpen
  \bibfield  {author} {\bibinfo {author} {\bibfnamefont {A.}~\bibnamefont
  {Nahum}}\ and\ \bibinfo {author} {\bibfnamefont {K.~J.}\ \bibnamefont
  {Wiese}},\ }\href {https://doi.org/10.48550/ARXIV.2303.07848} {}\Eprint
  {https://arxiv.org/abs/2303.07848} {arXiv:2303.07848} \BibitemShut {NoStop}%
\bibitem [{\citenamefont {Bao}\ \emph {et~al.}(2020)\citenamefont {Bao},
  \citenamefont {Choi},\ and\ \citenamefont {Altman}}]{bao2020theoryofthe}%
  \BibitemOpen
  \bibfield  {author} {\bibinfo {author} {\bibfnamefont {Y.}~\bibnamefont
  {Bao}}, \bibinfo {author} {\bibfnamefont {S.}~\bibnamefont {Choi}},\ and\
  \bibinfo {author} {\bibfnamefont {E.}~\bibnamefont {Altman}},\ }\href
  {https://doi.org/10.1103/PhysRevB.101.104301} {\bibfield  {journal} {\bibinfo
   {journal} {Phys. Rev. B}\ }\textbf {\bibinfo {volume} {101}},\ \bibinfo
  {pages} {104301} (\bibinfo {year} {2020})}\BibitemShut {NoStop}%
\bibitem [{\citenamefont {Choi}\ \emph {et~al.}(2020)\citenamefont {Choi},
  \citenamefont {Bao}, \citenamefont {Qi},\ and\ \citenamefont
  {Altman}}]{choi2020quantumerrorcorrection}%
  \BibitemOpen
  \bibfield  {author} {\bibinfo {author} {\bibfnamefont {S.}~\bibnamefont
  {Choi}}, \bibinfo {author} {\bibfnamefont {Y.}~\bibnamefont {Bao}}, \bibinfo
  {author} {\bibfnamefont {X.-L.}\ \bibnamefont {Qi}},\ and\ \bibinfo {author}
  {\bibfnamefont {E.}~\bibnamefont {Altman}},\ }\href
  {https://doi.org/10.1103/PhysRevLett.125.030505} {\bibfield  {journal}
  {\bibinfo  {journal} {Phys. Rev. Lett.}\ }\textbf {\bibinfo {volume} {125}},\
  \bibinfo {pages} {030505} (\bibinfo {year} {2020})}\BibitemShut {NoStop}%
\bibitem [{\citenamefont {Cao}\ \emph {et~al.}(2019)\citenamefont {Cao},
  \citenamefont {Tilloy},\ and\ \citenamefont {Luca}}]{cao2019entanglementina}%
  \BibitemOpen
  \bibfield  {author} {\bibinfo {author} {\bibfnamefont {X.}~\bibnamefont
  {Cao}}, \bibinfo {author} {\bibfnamefont {A.}~\bibnamefont {Tilloy}},\ and\
  \bibinfo {author} {\bibfnamefont {A.~D.}\ \bibnamefont {Luca}},\ }\href
  {https://doi.org/10.21468/SciPostPhys.7.2.024} {\bibfield  {journal}
  {\bibinfo  {journal} {SciPost Phys.}\ }\textbf {\bibinfo {volume} {7}},\
  \bibinfo {pages} {024} (\bibinfo {year} {2019})}\BibitemShut {NoStop}%
\bibitem [{\citenamefont {Alberton}\ \emph {et~al.}(2021)\citenamefont
  {Alberton}, \citenamefont {Buchhold},\ and\ \citenamefont
  {Diehl}}]{alberton2021entanglementtransitionin}%
  \BibitemOpen
  \bibfield  {author} {\bibinfo {author} {\bibfnamefont {O.}~\bibnamefont
  {Alberton}}, \bibinfo {author} {\bibfnamefont {M.}~\bibnamefont {Buchhold}},\
  and\ \bibinfo {author} {\bibfnamefont {S.}~\bibnamefont {Diehl}},\ }\href
  {https://doi.org/10.1103/PhysRevLett.126.170602} {\bibfield  {journal}
  {\bibinfo  {journal} {Phys. Rev. Lett.}\ }\textbf {\bibinfo {volume} {126}},\
  \bibinfo {pages} {170602} (\bibinfo {year} {2021})}\BibitemShut {NoStop}%
\bibitem [{\citenamefont {Botzung}\ \emph {et~al.}(2021)\citenamefont
  {Botzung}, \citenamefont {Diehl},\ and\ \citenamefont
  {M\"uller}}]{botzung2021engineereddissipationinduced}%
  \BibitemOpen
  \bibfield  {author} {\bibinfo {author} {\bibfnamefont {T.}~\bibnamefont
  {Botzung}}, \bibinfo {author} {\bibfnamefont {S.}~\bibnamefont {Diehl}},\
  and\ \bibinfo {author} {\bibfnamefont {M.}~\bibnamefont {M\"uller}},\ }\href
  {https://doi.org/10.1103/PhysRevB.104.184422} {\bibfield  {journal} {\bibinfo
   {journal} {Phys. Rev. B}\ }\textbf {\bibinfo {volume} {104}},\ \bibinfo
  {pages} {184422} (\bibinfo {year} {2021})}\BibitemShut {NoStop}%
\bibitem [{\citenamefont {Turkeshi}\ \emph
  {et~al.}(2022{\natexlab{a}})\citenamefont {Turkeshi}, \citenamefont
  {Piroli},\ and\ \citenamefont
  {Schir\'o}}]{turkeshi2022enhancedentanglementnegativity}%
  \BibitemOpen
  \bibfield  {author} {\bibinfo {author} {\bibfnamefont {X.}~\bibnamefont
  {Turkeshi}}, \bibinfo {author} {\bibfnamefont {L.}~\bibnamefont {Piroli}},\
  and\ \bibinfo {author} {\bibfnamefont {M.}~\bibnamefont {Schir\'o}},\ }\href
  {https://doi.org/10.1103/PhysRevB.106.024304} {\bibfield  {journal} {\bibinfo
   {journal} {Phys. Rev. B}\ }\textbf {\bibinfo {volume} {106}},\ \bibinfo
  {pages} {024304} (\bibinfo {year} {2022}{\natexlab{a}})}\BibitemShut
  {NoStop}%
\bibitem [{\citenamefont {Turkeshi}\ \emph {et~al.}(2021)\citenamefont
  {Turkeshi}, \citenamefont {Biella}, \citenamefont {Fazio}, \citenamefont
  {Dalmonte},\ and\ \citenamefont
  {Schir\'o}}]{turkeshi2021measurementinducedentanglement}%
  \BibitemOpen
  \bibfield  {author} {\bibinfo {author} {\bibfnamefont {X.}~\bibnamefont
  {Turkeshi}}, \bibinfo {author} {\bibfnamefont {A.}~\bibnamefont {Biella}},
  \bibinfo {author} {\bibfnamefont {R.}~\bibnamefont {Fazio}}, \bibinfo
  {author} {\bibfnamefont {M.}~\bibnamefont {Dalmonte}},\ and\ \bibinfo
  {author} {\bibfnamefont {M.}~\bibnamefont {Schir\'o}},\ }\href
  {https://doi.org/10.1103/PhysRevB.103.224210} {\bibfield  {journal} {\bibinfo
   {journal} {Phys. Rev. B}\ }\textbf {\bibinfo {volume} {103}},\ \bibinfo
  {pages} {224210} (\bibinfo {year} {2021})}\BibitemShut {NoStop}%
\bibitem [{\citenamefont {Buchhold}\ \emph {et~al.}()\citenamefont {Buchhold},
  \citenamefont {M\"ller},\ and\ \citenamefont
  {Diehl}}]{buchhold2022revealingmeasurementinduced}%
  \BibitemOpen
  \bibfield  {author} {\bibinfo {author} {\bibfnamefont {M.}~\bibnamefont
  {Buchhold}}, \bibinfo {author} {\bibfnamefont {T.}~\bibnamefont {M\"ller}},\
  and\ \bibinfo {author} {\bibfnamefont {S.}~\bibnamefont {Diehl}},\ }\href
  {https://doi.org/10.48550/ARXIV.2208.10506} {}\Eprint
  {https://arxiv.org/abs/2208.10506} {arXiv:2208.10506} \BibitemShut {NoStop}%
\bibitem [{\citenamefont {Minato}\ \emph {et~al.}(2022)\citenamefont {Minato},
  \citenamefont {Sugimoto}, \citenamefont {Kuwahara},\ and\ \citenamefont
  {Saito}}]{minato2022fateofmeasurementinduced}%
  \BibitemOpen
  \bibfield  {author} {\bibinfo {author} {\bibfnamefont {T.}~\bibnamefont
  {Minato}}, \bibinfo {author} {\bibfnamefont {K.}~\bibnamefont {Sugimoto}},
  \bibinfo {author} {\bibfnamefont {T.}~\bibnamefont {Kuwahara}},\ and\
  \bibinfo {author} {\bibfnamefont {K.}~\bibnamefont {Saito}},\ }\href
  {https://doi.org/10.1103/PhysRevLett.128.010603} {\bibfield  {journal}
  {\bibinfo  {journal} {Phys. Rev. Lett.}\ }\textbf {\bibinfo {volume} {128}},\
  \bibinfo {pages} {010603} (\bibinfo {year} {2022})}\BibitemShut {NoStop}%
\bibitem [{\citenamefont {Ladewig}\ \emph {et~al.}(2022)\citenamefont
  {Ladewig}, \citenamefont {Diehl},\ and\ \citenamefont
  {Buchhold}}]{ladewig2022monitoredopenfermion}%
  \BibitemOpen
  \bibfield  {author} {\bibinfo {author} {\bibfnamefont {B.}~\bibnamefont
  {Ladewig}}, \bibinfo {author} {\bibfnamefont {S.}~\bibnamefont {Diehl}},\
  and\ \bibinfo {author} {\bibfnamefont {M.}~\bibnamefont {Buchhold}},\ }\href
  {https://doi.org/10.1103/PhysRevResearch.4.033001} {\bibfield  {journal}
  {\bibinfo  {journal} {Phys. Rev. Res.}\ }\textbf {\bibinfo {volume} {4}},\
  \bibinfo {pages} {033001} (\bibinfo {year} {2022})}\BibitemShut {NoStop}%
\bibitem [{\citenamefont {Piccitto}\ \emph {et~al.}(2022)\citenamefont
  {Piccitto}, \citenamefont {Russomanno},\ and\ \citenamefont
  {Rossini}}]{piccitto2022entanglementtransitionsin}%
  \BibitemOpen
  \bibfield  {author} {\bibinfo {author} {\bibfnamefont {G.}~\bibnamefont
  {Piccitto}}, \bibinfo {author} {\bibfnamefont {A.}~\bibnamefont
  {Russomanno}},\ and\ \bibinfo {author} {\bibfnamefont {D.}~\bibnamefont
  {Rossini}},\ }\href {https://doi.org/10.1103/PhysRevB.105.064305} {\bibfield
  {journal} {\bibinfo  {journal} {Phys. Rev. B}\ }\textbf {\bibinfo {volume}
  {105}},\ \bibinfo {pages} {064305} (\bibinfo {year} {2022})}\BibitemShut
  {NoStop}%
\bibitem [{\citenamefont {Gal}\ \emph {et~al.}()\citenamefont {Gal},
  \citenamefont {Turkeshi},\ and\ \citenamefont
  {Schirò}}]{gal2022volumetoarea}%
  \BibitemOpen
  \bibfield  {author} {\bibinfo {author} {\bibfnamefont {Y.~L.}\ \bibnamefont
  {Gal}}, \bibinfo {author} {\bibfnamefont {X.}~\bibnamefont {Turkeshi}},\ and\
  \bibinfo {author} {\bibfnamefont {M.}~\bibnamefont {Schirò}},\ }\href@noop
  {} {}\Eprint {https://arxiv.org/abs/2210.11937} {arXiv:2210.11937}
  \BibitemShut {NoStop}%
\bibitem [{\citenamefont {Granet}\ \emph {et~al.}()\citenamefont {Granet},
  \citenamefont {Zhang},\ and\ \citenamefont
  {Dreyer}}]{granet2022volumelawtoarealaw}%
  \BibitemOpen
  \bibfield  {author} {\bibinfo {author} {\bibfnamefont {E.}~\bibnamefont
  {Granet}}, \bibinfo {author} {\bibfnamefont {C.}~\bibnamefont {Zhang}},\ and\
  \bibinfo {author} {\bibfnamefont {H.}~\bibnamefont {Dreyer}},\ }\href
  {https://doi.org/10.48550/ARXIV.2212.10584} {}\Eprint
  {https://arxiv.org/abs/2212.10584} {arXiv:2212.10584} \BibitemShut {NoStop}%
\bibitem [{\citenamefont {M\"uller}\ \emph {et~al.}(2022)\citenamefont
  {M\"uller}, \citenamefont {Diehl},\ and\ \citenamefont
  {Buchhold}}]{muller2022measurementinduceddark}%
  \BibitemOpen
  \bibfield  {author} {\bibinfo {author} {\bibfnamefont {T.}~\bibnamefont
  {M\"uller}}, \bibinfo {author} {\bibfnamefont {S.}~\bibnamefont {Diehl}},\
  and\ \bibinfo {author} {\bibfnamefont {M.}~\bibnamefont {Buchhold}},\ }\href
  {https://doi.org/10.1103/PhysRevLett.128.010605} {\bibfield  {journal}
  {\bibinfo  {journal} {Phys. Rev. Lett.}\ }\textbf {\bibinfo {volume} {128}},\
  \bibinfo {pages} {010605} (\bibinfo {year} {2022})}\BibitemShut {NoStop}%
\bibitem [{\citenamefont {Buchhold}\ \emph {et~al.}(2021)\citenamefont
  {Buchhold}, \citenamefont {Minoguchi}, \citenamefont {Altland},\ and\
  \citenamefont {Diehl}}]{buchhold2021effectivetheoryfor}%
  \BibitemOpen
  \bibfield  {author} {\bibinfo {author} {\bibfnamefont {M.}~\bibnamefont
  {Buchhold}}, \bibinfo {author} {\bibfnamefont {Y.}~\bibnamefont {Minoguchi}},
  \bibinfo {author} {\bibfnamefont {A.}~\bibnamefont {Altland}},\ and\ \bibinfo
  {author} {\bibfnamefont {S.}~\bibnamefont {Diehl}},\ }\href
  {https://doi.org/10.1103/PhysRevX.11.041004} {\bibfield  {journal} {\bibinfo
  {journal} {Phys. Rev. X}\ }\textbf {\bibinfo {volume} {11}},\ \bibinfo
  {pages} {041004} (\bibinfo {year} {2021})}\BibitemShut {NoStop}%
\bibitem [{\citenamefont {Turkeshi}\ \emph
  {et~al.}(2022{\natexlab{b}})\citenamefont {Turkeshi}, \citenamefont
  {Dalmonte}, \citenamefont {Fazio},\ and\ \citenamefont
  {Schir\`o}}]{turkeshi2022entanglementtransitionsfrom}%
  \BibitemOpen
  \bibfield  {author} {\bibinfo {author} {\bibfnamefont {X.}~\bibnamefont
  {Turkeshi}}, \bibinfo {author} {\bibfnamefont {M.}~\bibnamefont {Dalmonte}},
  \bibinfo {author} {\bibfnamefont {R.}~\bibnamefont {Fazio}},\ and\ \bibinfo
  {author} {\bibfnamefont {M.}~\bibnamefont {Schir\`o}},\ }\href
  {https://doi.org/10.1103/PhysRevB.105.L241114} {\bibfield  {journal}
  {\bibinfo  {journal} {Phys. Rev. B}\ }\textbf {\bibinfo {volume} {105}},\
  \bibinfo {pages} {L241114} (\bibinfo {year}
  {2022}{\natexlab{b}})}\BibitemShut {NoStop}%
\bibitem [{\citenamefont {Biella}\ and\ \citenamefont
  {Schir{\'{o}}}(2021)}]{biella2021manybodyquantumzeno}%
  \BibitemOpen
  \bibfield  {author} {\bibinfo {author} {\bibfnamefont {A.}~\bibnamefont
  {Biella}}\ and\ \bibinfo {author} {\bibfnamefont {M.}~\bibnamefont
  {Schir{\'{o}}}},\ }\href {https://doi.org/10.22331/q-2021-08-19-528}
  {\bibfield  {journal} {\bibinfo  {journal} {{Quantum}}\ }\textbf {\bibinfo
  {volume} {5}},\ \bibinfo {pages} {528} (\bibinfo {year} {2021})}\BibitemShut
  {NoStop}%
\bibitem [{\citenamefont {Turkeshi}\ and\ \citenamefont
  {Schir\'o}(2023)}]{turkeshi2022entanglementandcorrelation}%
  \BibitemOpen
  \bibfield  {author} {\bibinfo {author} {\bibfnamefont {X.}~\bibnamefont
  {Turkeshi}}\ and\ \bibinfo {author} {\bibfnamefont {M.}~\bibnamefont
  {Schir\'o}},\ }\href {https://doi.org/10.1103/PhysRevB.107.L020403}
  {\bibfield  {journal} {\bibinfo  {journal} {Phys. Rev. B}\ }\textbf {\bibinfo
  {volume} {107}},\ \bibinfo {pages} {L020403} (\bibinfo {year}
  {2023})}\BibitemShut {NoStop}%
\bibitem [{\citenamefont {Nahum}\ and\ \citenamefont
  {Skinner}(2020)}]{nahum2020entanglementanddynamics}%
  \BibitemOpen
  \bibfield  {author} {\bibinfo {author} {\bibfnamefont {A.}~\bibnamefont
  {Nahum}}\ and\ \bibinfo {author} {\bibfnamefont {B.}~\bibnamefont
  {Skinner}},\ }\href {https://doi.org/10.1103/PhysRevResearch.2.023288}
  {\bibfield  {journal} {\bibinfo  {journal} {Phys. Rev. Res.}\ }\textbf
  {\bibinfo {volume} {2}},\ \bibinfo {pages} {023288} (\bibinfo {year}
  {2020})}\BibitemShut {NoStop}%
\bibitem [{\citenamefont {Fidkowski}\ \emph
  {et~al.}(2021{\natexlab{a}})\citenamefont {Fidkowski}, \citenamefont {Haah},\
  and\ \citenamefont {Hastings}}]{fidkowski2021howdynamicalquantum}%
  \BibitemOpen
  \bibfield  {author} {\bibinfo {author} {\bibfnamefont {L.}~\bibnamefont
  {Fidkowski}}, \bibinfo {author} {\bibfnamefont {J.}~\bibnamefont {Haah}},\
  and\ \bibinfo {author} {\bibfnamefont {M.~B.}\ \bibnamefont {Hastings}},\
  }\href {https://doi.org/10.22331/q-2021-01-17-382} {\bibfield  {journal}
  {\bibinfo  {journal} {{Quantum}}\ }\textbf {\bibinfo {volume} {5}},\ \bibinfo
  {pages} {382} (\bibinfo {year} {2021}{\natexlab{a}})}\BibitemShut {NoStop}%
\bibitem [{\citenamefont {Fidkowski}\ \emph
  {et~al.}(2021{\natexlab{b}})\citenamefont {Fidkowski}, \citenamefont {Haah},\
  and\ \citenamefont {Hastings}}]{fidkowski2021dynamical}%
  \BibitemOpen
  \bibfield  {author} {\bibinfo {author} {\bibfnamefont {L.}~\bibnamefont
  {Fidkowski}}, \bibinfo {author} {\bibfnamefont {J.}~\bibnamefont {Haah}},\
  and\ \bibinfo {author} {\bibfnamefont {M.~B.}\ \bibnamefont {Hastings}},\
  }\href {https://doi.org/10.22331/q-2021-01-17-382} {\bibfield  {journal}
  {\bibinfo  {journal} {Quantum}\ }\textbf {\bibinfo {volume} {5}},\ \bibinfo
  {pages} {382} (\bibinfo {year} {2021}{\natexlab{b}})}\BibitemShut {NoStop}%
\bibitem [{\citenamefont {Fava}\ \emph {et~al.}()\citenamefont {Fava},
  \citenamefont {Piroli}, \citenamefont {Swann}, \citenamefont {Bernard},\ and\
  \citenamefont {Nahum}}]{fava2023nonlinearsigmamodels}%
  \BibitemOpen
  \bibfield  {author} {\bibinfo {author} {\bibfnamefont {M.}~\bibnamefont
  {Fava}}, \bibinfo {author} {\bibfnamefont {L.}~\bibnamefont {Piroli}},
  \bibinfo {author} {\bibfnamefont {T.}~\bibnamefont {Swann}}, \bibinfo
  {author} {\bibfnamefont {D.}~\bibnamefont {Bernard}},\ and\ \bibinfo {author}
  {\bibfnamefont {A.}~\bibnamefont {Nahum}},\ }\href
  {https://doi.org/10.48550/ARXIV.2302.12820} {}\Eprint
  {https://arxiv.org/abs/2302.12820} {arXiv:2302.12820} \BibitemShut {NoStop}%
\bibitem [{\citenamefont {Bao}\ \emph {et~al.}(2021)\citenamefont {Bao},
  \citenamefont {Choi},\ and\ \citenamefont
  {Altman}}]{bao2021symmetryenrichedphases}%
  \BibitemOpen
  \bibfield  {author} {\bibinfo {author} {\bibfnamefont {Y.}~\bibnamefont
  {Bao}}, \bibinfo {author} {\bibfnamefont {S.}~\bibnamefont {Choi}},\ and\
  \bibinfo {author} {\bibfnamefont {E.}~\bibnamefont {Altman}},\ }\href
  {https://doi.org/10.1016/j.aop.2021.168618} {\bibfield  {journal} {\bibinfo
  {journal} {Ann. Phys.}\ }\textbf {\bibinfo {volume} {435}},\ \bibinfo {pages}
  {168618} (\bibinfo {year} {2021})}\BibitemShut {NoStop}%
\bibitem [{\citenamefont {Jian}\ \emph {et~al.}(2022)\citenamefont {Jian},
  \citenamefont {Bauer}, \citenamefont {Keselman},\ and\ \citenamefont
  {Ludwig}}]{chaoming2022criticalityandentanglement}%
  \BibitemOpen
  \bibfield  {author} {\bibinfo {author} {\bibfnamefont {C.-M.}\ \bibnamefont
  {Jian}}, \bibinfo {author} {\bibfnamefont {B.}~\bibnamefont {Bauer}},
  \bibinfo {author} {\bibfnamefont {A.}~\bibnamefont {Keselman}},\ and\
  \bibinfo {author} {\bibfnamefont {A.~W.~W.}\ \bibnamefont {Ludwig}},\ }\href
  {https://doi.org/10.1103/PhysRevB.106.134206} {\bibfield  {journal} {\bibinfo
   {journal} {Phys. Rev. B}\ }\textbf {\bibinfo {volume} {106}},\ \bibinfo
  {pages} {134206} (\bibinfo {year} {2022})}\BibitemShut {NoStop}%
\bibitem [{\citenamefont {{C.-M. Jian, H. Shapourian, B. Bauer, and A. W. W.
  Ludwig}}()}]{chaoming2023measurementinducedentanglementtransitions}%
  \BibitemOpen
  \bibfield  {author} {\bibinfo {author} {\bibnamefont {{C.-M. Jian, H.
  Shapourian, B. Bauer, and A. W. W. Ludwig}}},\ }\href
  {https://doi.org/10.48550/ARXIV.2302.09094} {}\Eprint
  {https://arxiv.org/abs/2302.09094} {arXiv:2302.09094} \BibitemShut {NoStop}%
\bibitem [{\citenamefont {Merritt}\ and\ \citenamefont
  {Fidkowski}(2023)}]{merritt2023entanglementtransitionswith}%
  \BibitemOpen
  \bibfield  {author} {\bibinfo {author} {\bibfnamefont {J.}~\bibnamefont
  {Merritt}}\ and\ \bibinfo {author} {\bibfnamefont {L.}~\bibnamefont
  {Fidkowski}},\ }\href {https://doi.org/10.1103/PhysRevB.107.064303}
  {\bibfield  {journal} {\bibinfo  {journal} {Phys. Rev. B}\ }\textbf {\bibinfo
  {volume} {107}},\ \bibinfo {pages} {064303} (\bibinfo {year}
  {2023})}\BibitemShut {NoStop}%
\bibitem [{\citenamefont {Bocquet}\ \emph {et~al.}(2000)\citenamefont
  {Bocquet}, \citenamefont {Serban},\ and\ \citenamefont
  {Zirnbauer}}]{bocquet2000disordered2dquasiparticles}%
  \BibitemOpen
  \bibfield  {author} {\bibinfo {author} {\bibfnamefont {M.}~\bibnamefont
  {Bocquet}}, \bibinfo {author} {\bibfnamefont {D.}~\bibnamefont {Serban}},\
  and\ \bibinfo {author} {\bibfnamefont {M.}~\bibnamefont {Zirnbauer}},\ }\href
  {https://doi.org/10.1016/s0550-3213(00)00208-x} {\bibfield  {journal}
  {\bibinfo  {journal} {Nucl. Phys. B}\ }\textbf {\bibinfo {volume} {578}},\
  \bibinfo {pages} {628} (\bibinfo {year} {2000})}\BibitemShut {NoStop}%
\bibitem [{\citenamefont {Senthil}\ and\ \citenamefont
  {Fisher}(2000)}]{senthil2000quasiparticlelocalizationin}%
  \BibitemOpen
  \bibfield  {author} {\bibinfo {author} {\bibfnamefont {T.}~\bibnamefont
  {Senthil}}\ and\ \bibinfo {author} {\bibfnamefont {M.~P.~A.}\ \bibnamefont
  {Fisher}},\ }\href {https://doi.org/10.1103/PhysRevB.61.9690} {\bibfield
  {journal} {\bibinfo  {journal} {Phys. Rev. B}\ }\textbf {\bibinfo {volume}
  {61}},\ \bibinfo {pages} {9690} (\bibinfo {year} {2000})}\BibitemShut
  {NoStop}%
\bibitem [{\citenamefont {Chalker}\ \emph {et~al.}(2001)\citenamefont
  {Chalker}, \citenamefont {Read}, \citenamefont {Kagalovsky}, \citenamefont
  {Horovitz}, \citenamefont {Avishai},\ and\ \citenamefont
  {Ludwig}}]{chalker2001thermalmetalin}%
  \BibitemOpen
  \bibfield  {author} {\bibinfo {author} {\bibfnamefont {J.~T.}\ \bibnamefont
  {Chalker}}, \bibinfo {author} {\bibfnamefont {N.}~\bibnamefont {Read}},
  \bibinfo {author} {\bibfnamefont {V.}~\bibnamefont {Kagalovsky}}, \bibinfo
  {author} {\bibfnamefont {B.}~\bibnamefont {Horovitz}}, \bibinfo {author}
  {\bibfnamefont {Y.}~\bibnamefont {Avishai}},\ and\ \bibinfo {author}
  {\bibfnamefont {A.~W.~W.}\ \bibnamefont {Ludwig}},\ }\href
  {https://doi.org/10.1103/PhysRevB.65.012506} {\bibfield  {journal} {\bibinfo
  {journal} {Phys. Rev. B}\ }\textbf {\bibinfo {volume} {65}},\ \bibinfo
  {pages} {012506} (\bibinfo {year} {2001})}\BibitemShut {NoStop}%
\bibitem [{\citenamefont {Evers}\ and\ \citenamefont
  {Mirlin}(2008)}]{evers2008andersontransitions}%
  \BibitemOpen
  \bibfield  {author} {\bibinfo {author} {\bibfnamefont {F.}~\bibnamefont
  {Evers}}\ and\ \bibinfo {author} {\bibfnamefont {A.~D.}\ \bibnamefont
  {Mirlin}},\ }\href {https://doi.org/10.1103/RevModPhys.80.1355} {\bibfield
  {journal} {\bibinfo  {journal} {Rev. Mod. Phys.}\ }\textbf {\bibinfo {volume}
  {80}},\ \bibinfo {pages} {1355} (\bibinfo {year} {2008})}\BibitemShut
  {NoStop}%
\bibitem [{\citenamefont {Amico}\ \emph {et~al.}(2008)\citenamefont {Amico},
  \citenamefont {Fazio}, \citenamefont {Osterloh},\ and\ \citenamefont
  {Vedral}}]{amico2008entanglementinmanybody}%
  \BibitemOpen
  \bibfield  {author} {\bibinfo {author} {\bibfnamefont {L.}~\bibnamefont
  {Amico}}, \bibinfo {author} {\bibfnamefont {R.}~\bibnamefont {Fazio}},
  \bibinfo {author} {\bibfnamefont {A.}~\bibnamefont {Osterloh}},\ and\
  \bibinfo {author} {\bibfnamefont {V.}~\bibnamefont {Vedral}},\ }\href
  {https://doi.org/10.1103/RevModPhys.80.517} {\bibfield  {journal} {\bibinfo
  {journal} {Rev. Mod. Phys.}\ }\textbf {\bibinfo {volume} {80}},\ \bibinfo
  {pages} {517} (\bibinfo {year} {2008})}\BibitemShut {NoStop}%
\bibitem [{\citenamefont {Calabrese}\ and\ \citenamefont
  {Cardy}(2004)}]{Calabrese_2004}%
  \BibitemOpen
  \bibfield  {author} {\bibinfo {author} {\bibfnamefont {P.}~\bibnamefont
  {Calabrese}}\ and\ \bibinfo {author} {\bibfnamefont {J.}~\bibnamefont
  {Cardy}},\ }\href {https://doi.org/10.1088/1742-5468/2004/06/p06002}
  {\bibfield  {journal} {\bibinfo  {journal} {J. Stat. Mech.}\ }\textbf
  {\bibinfo {volume} {2004}},\ \bibinfo {pages} {P06002} (\bibinfo {year}
  {2004})}\BibitemShut {NoStop}%
\bibitem [{\citenamefont {Calabrese}\ and\ \citenamefont
  {Cardy}(2009)}]{Calabrese_2009}%
  \BibitemOpen
  \bibfield  {author} {\bibinfo {author} {\bibfnamefont {P.}~\bibnamefont
  {Calabrese}}\ and\ \bibinfo {author} {\bibfnamefont {J.}~\bibnamefont
  {Cardy}},\ }\href {https://doi.org/10.1088/1751-8113/42/50/504005} {\bibfield
   {journal} {\bibinfo  {journal} {J. Phys. A}\ }\textbf {\bibinfo {volume}
  {42}},\ \bibinfo {pages} {504005} (\bibinfo {year} {2009})}\BibitemShut
  {NoStop}%
\bibitem [{\citenamefont {Block}\ \emph {et~al.}(2022)\citenamefont {Block},
  \citenamefont {Bao}, \citenamefont {Choi}, \citenamefont {Altman},\ and\
  \citenamefont {Yao}}]{block2022measurementinducedtransition}%
  \BibitemOpen
  \bibfield  {author} {\bibinfo {author} {\bibfnamefont {M.}~\bibnamefont
  {Block}}, \bibinfo {author} {\bibfnamefont {Y.}~\bibnamefont {Bao}}, \bibinfo
  {author} {\bibfnamefont {S.}~\bibnamefont {Choi}}, \bibinfo {author}
  {\bibfnamefont {E.}~\bibnamefont {Altman}},\ and\ \bibinfo {author}
  {\bibfnamefont {N.~Y.}\ \bibnamefont {Yao}},\ }\href
  {https://doi.org/10.1103/PhysRevLett.128.010604} {\bibfield  {journal}
  {\bibinfo  {journal} {Phys. Rev. Lett.}\ }\textbf {\bibinfo {volume} {128}},\
  \bibinfo {pages} {010604} (\bibinfo {year} {2022})}\BibitemShut {NoStop}%
\bibitem [{\citenamefont {Sharma}\ \emph {et~al.}(2022)\citenamefont {Sharma},
  \citenamefont {Turkeshi}, \citenamefont {Fazio},\ and\ \citenamefont
  {Dalmonte}}]{sharma2022measurementinducedcriticality}%
  \BibitemOpen
  \bibfield  {author} {\bibinfo {author} {\bibfnamefont {S.}~\bibnamefont
  {Sharma}}, \bibinfo {author} {\bibfnamefont {X.}~\bibnamefont {Turkeshi}},
  \bibinfo {author} {\bibfnamefont {R.}~\bibnamefont {Fazio}},\ and\ \bibinfo
  {author} {\bibfnamefont {M.}~\bibnamefont {Dalmonte}},\ }\href
  {https://doi.org/10.21468/SciPostPhysCore.5.2.023} {\bibfield  {journal}
  {\bibinfo  {journal} {SciPost Phys. Core}\ }\textbf {\bibinfo {volume} {5}},\
  \bibinfo {pages} {023} (\bibinfo {year} {2022})}\BibitemShut {NoStop}%
\bibitem [{\citenamefont {Koh}\ \emph {et~al.}()\citenamefont {Koh},
  \citenamefont {Sun}, \citenamefont {Motta},\ and\ \citenamefont
  {Minnich}}]{koh2022experimentalrealizationof}%
  \BibitemOpen
  \bibfield  {author} {\bibinfo {author} {\bibfnamefont {J.~M.}\ \bibnamefont
  {Koh}}, \bibinfo {author} {\bibfnamefont {S.-N.}\ \bibnamefont {Sun}},
  \bibinfo {author} {\bibfnamefont {M.}~\bibnamefont {Motta}},\ and\ \bibinfo
  {author} {\bibfnamefont {A.~J.}\ \bibnamefont {Minnich}},\ }\href@noop {}
  {}\Eprint {https://arxiv.org/abs/2203.04338} {arXiv:2203.04338} \BibitemShut
  {NoStop}%
\bibitem [{\citenamefont {Noel}\ \emph {et~al.}(2022)\citenamefont {Noel},
  \citenamefont {Niroula}, \citenamefont {Zhu}, \citenamefont {Risinger},
  \citenamefont {Egan}, \citenamefont {Biswas}, \citenamefont {Cetina},
  \citenamefont {Gorshkov}, \citenamefont {Gullans}, \citenamefont {Huse},\
  and\ \citenamefont {Monroe}}]{noel2022measurementinducedquantum}%
  \BibitemOpen
  \bibfield  {author} {\bibinfo {author} {\bibfnamefont {C.}~\bibnamefont
  {Noel}}, \bibinfo {author} {\bibfnamefont {P.}~\bibnamefont {Niroula}},
  \bibinfo {author} {\bibfnamefont {D.}~\bibnamefont {Zhu}}, \bibinfo {author}
  {\bibfnamefont {A.}~\bibnamefont {Risinger}}, \bibinfo {author}
  {\bibfnamefont {L.}~\bibnamefont {Egan}}, \bibinfo {author} {\bibfnamefont
  {D.}~\bibnamefont {Biswas}}, \bibinfo {author} {\bibfnamefont
  {M.}~\bibnamefont {Cetina}}, \bibinfo {author} {\bibfnamefont {A.~V.}\
  \bibnamefont {Gorshkov}}, \bibinfo {author} {\bibfnamefont {M.~J.}\
  \bibnamefont {Gullans}}, \bibinfo {author} {\bibfnamefont {D.~A.}\
  \bibnamefont {Huse}},\ and\ \bibinfo {author} {\bibfnamefont
  {C.}~\bibnamefont {Monroe}},\ }\href
  {https://doi.org/10.1038/s41567-022-01619-7} {\bibfield  {journal} {\bibinfo
  {journal} {Nat. Phys.}\ }\textbf {\bibinfo {volume} {18}},\ \bibinfo {pages}
  {760} (\bibinfo {year} {2022})}\BibitemShut {NoStop}%
\bibitem [{\citenamefont
  {Bravyi}(2005)}]{bravyi2004lagrangianrepresentationfor}%
  \BibitemOpen
  \bibfield  {author} {\bibinfo {author} {\bibfnamefont {S.}~\bibnamefont
  {Bravyi}},\ }\href {https://doi.org/10.48550/ARXIV.QUANT-PH/0404180}
  {\bibfield  {journal} {\bibinfo  {journal} {Quantum Inf. and Comp.}\ }\textbf
  {\bibinfo {volume} {5}},\ \bibinfo {pages} {216} (\bibinfo {year}
  {2005})}\BibitemShut {NoStop}%
\bibitem [{sup()}]{supmat}%
  \BibitemOpen
  \href@noop {} {\bibinfo {title} {See the supplementary material, including
  ref.~\cite{Kawashima1993}, for details on the numerical implementation and
  additional numerical data.}}\BibitemShut {Stop}%
\bibitem [{Note1()}]{Note1}%
  \BibitemOpen
  \bibinfo {note} {Indeed, in this case, the initial system-ancilla state is
  pure, $S_2(\rho _S(t))=S_2(\rho _A(t))$ allowing considerable simplification
  in the numerical calculations.}\BibitemShut {Stop}%
\bibitem [{Note2()}]{Note2}%
  \BibitemOpen
  \bibinfo {note} {\protect \textit {En passant}, we note that considering
  other indicators such as the average value give qualitatively similar
  results.}\BibitemShut {Stop}%
\bibitem [{Note3()}]{Note3}%
  \BibitemOpen
  \bibinfo {note} {To account for the logarithmic behavior for high rates
  $p\simeq 1$, we consider $f(x) = \beta (L^\alpha -1)/\alpha $. \protect
  \textit {En passant}, we note that the available data cannot exclude the
  presence of multiplicative logarithms, arising in analogy to $\protect
  \mathbb {Z}_2$ case, cf.~\cite {supmat}.}\BibitemShut {Stop}%
\bibitem [{\citenamefont {{M. Bucchold, et al.}}()}]{buccholdtoappear}%
  \BibitemOpen
  \bibfield  {author} {\bibinfo {author} {\bibnamefont {{M. Bucchold, et
  al.}}},\ }\href@noop {} {\bibinfo {title} {To appear,}}\BibitemShut {NoStop}%
\bibitem [{\citenamefont {Ho}\ and\ \citenamefont
  {Choi}(2022)}]{ho2022exactemergentquantum}%
  \BibitemOpen
  \bibfield  {author} {\bibinfo {author} {\bibfnamefont {W.~W.}\ \bibnamefont
  {Ho}}\ and\ \bibinfo {author} {\bibfnamefont {S.}~\bibnamefont {Choi}},\
  }\href {https://doi.org/10.1103/PhysRevLett.128.060601} {\bibfield  {journal}
  {\bibinfo  {journal} {Phys. Rev. Lett.}\ }\textbf {\bibinfo {volume} {128}},\
  \bibinfo {pages} {060601} (\bibinfo {year} {2022})}\BibitemShut {NoStop}%
\bibitem [{\citenamefont {Claeys}\ and\ \citenamefont
  {Lamacraft}(2022)}]{Claeys2022emergentquantum}%
  \BibitemOpen
  \bibfield  {author} {\bibinfo {author} {\bibfnamefont {P.~W.}\ \bibnamefont
  {Claeys}}\ and\ \bibinfo {author} {\bibfnamefont {A.}~\bibnamefont
  {Lamacraft}},\ }\href {https://doi.org/10.22331/q-2022-06-15-738} {\bibfield
  {journal} {\bibinfo  {journal} {{Quantum}}\ }\textbf {\bibinfo {volume}
  {6}},\ \bibinfo {pages} {738} (\bibinfo {year} {2022})}\BibitemShut {NoStop}%
\bibitem [{\citenamefont {Ippoliti}\ and\ \citenamefont
  {Ho}(2022)}]{Ippoliti2022solvablemodelofdeep}%
  \BibitemOpen
  \bibfield  {author} {\bibinfo {author} {\bibfnamefont {M.}~\bibnamefont
  {Ippoliti}}\ and\ \bibinfo {author} {\bibfnamefont {W.~W.}\ \bibnamefont
  {Ho}},\ }\href {https://doi.org/10.22331/q-2022-12-29-886} {\bibfield
  {journal} {\bibinfo  {journal} {{Quantum}}\ }\textbf {\bibinfo {volume}
  {6}},\ \bibinfo {pages} {886} (\bibinfo {year} {2022})}\BibitemShut {NoStop}%
\bibitem [{\citenamefont {{M. Ippoliti and W. W.
  Ho}}()}]{ippoliti2022dynamicalpurificationand}%
  \BibitemOpen
  \bibfield  {author} {\bibinfo {author} {\bibnamefont {{M. Ippoliti and W. W.
  Ho}}},\ }\href {https://doi.org/10.48550/ARXIV.2204.13657} {}\Eprint
  {https://arxiv.org/abs/2204.13657} {arXiv:2204.13657} \BibitemShut {NoStop}%
\bibitem [{\citenamefont {Lucas}\ \emph {et~al.}()\citenamefont {Lucas},
  \citenamefont {Piroli}, \citenamefont {De~Nardis},\ and\ \citenamefont
  {De~Luca}}]{lucas2022generalizeddeepthermalization}%
  \BibitemOpen
  \bibfield  {author} {\bibinfo {author} {\bibfnamefont {M.}~\bibnamefont
  {Lucas}}, \bibinfo {author} {\bibfnamefont {L.}~\bibnamefont {Piroli}},
  \bibinfo {author} {\bibfnamefont {J.}~\bibnamefont {De~Nardis}},\ and\
  \bibinfo {author} {\bibfnamefont {A.}~\bibnamefont {De~Luca}},\ }\href
  {https://doi.org/10.48550/ARXIV.2207.13628} {}\Eprint
  {https://arxiv.org/abs/2207.13628} {arXiv:2207.13628} \BibitemShut {NoStop}%
\bibitem [{\citenamefont {\L{}ydzba}\ \emph {et~al.}()\citenamefont
  {\L{}ydzba}, \citenamefont {Mierzejewski}, \citenamefont {Rigol},\ and\
  \citenamefont {Vidmar}}]{lydzba2022}%
  \BibitemOpen
  \bibfield  {author} {\bibinfo {author} {\bibfnamefont {P.}~\bibnamefont
  {\L{}ydzba}}, \bibinfo {author} {\bibfnamefont {M.}~\bibnamefont
  {Mierzejewski}}, \bibinfo {author} {\bibfnamefont {M.}~\bibnamefont
  {Rigol}},\ and\ \bibinfo {author} {\bibfnamefont {L.}~\bibnamefont
  {Vidmar}},\ }\href {https://doi.org/10.48550/ARXIV.2210.00016} {}\Eprint
  {https://arxiv.org/abs/2210.00016} {arXiv:2210.00016} \BibitemShut {NoStop}%
\bibitem [{\citenamefont {Coppola}\ \emph {et~al.}(2022)\citenamefont
  {Coppola}, \citenamefont {Tirrito}, \citenamefont {Karevski},\ and\
  \citenamefont {Collura}}]{coppola2022growth}%
  \BibitemOpen
  \bibfield  {author} {\bibinfo {author} {\bibfnamefont {M.}~\bibnamefont
  {Coppola}}, \bibinfo {author} {\bibfnamefont {E.}~\bibnamefont {Tirrito}},
  \bibinfo {author} {\bibfnamefont {D.}~\bibnamefont {Karevski}},\ and\
  \bibinfo {author} {\bibfnamefont {M.}~\bibnamefont {Collura}},\ }\href
  {https://doi.org/10.1103/PhysRevB.105.094303} {\bibfield  {journal} {\bibinfo
   {journal} {Phys. Rev. B}\ }\textbf {\bibinfo {volume} {105}},\ \bibinfo
  {pages} {094303} (\bibinfo {year} {2022})}\BibitemShut {NoStop}%
\bibitem [{\citenamefont {Kawashima}\ and\ \citenamefont
  {Ito}(1993)}]{Kawashima1993}%
  \BibitemOpen
  \bibfield  {author} {\bibinfo {author} {\bibfnamefont {N.}~\bibnamefont
  {Kawashima}}\ and\ \bibinfo {author} {\bibfnamefont {N.}~\bibnamefont
  {Ito}},\ }\href {https://doi.org/10.1143/jpsj.62.435} {\bibfield  {journal}
  {\bibinfo  {journal} {J. Phys. Soc. Japan}\ }\textbf {\bibinfo {volume}
  {62}},\ \bibinfo {pages} {435} (\bibinfo {year} {1993})}\BibitemShut
  {NoStop}%
\end{thebibliography}%
\bibliographystyle{apsrev4-2}

\onecolumngrid
\newpage

\appendix

\begin{center}
{\Large Supplementary Material \\ 
\titleinfo
}
\end{center}

\setcounter{equation}{0}
\setcounter{figure}{0}
\setcounter{table}{0}
\setcounter{page}{1}
\renewcommand{\theequation}{S\arabic{equation}}
\setcounter{figure}{0}
\renewcommand{\thefigure}{S\arabic{figure}}
\renewcommand{\thepage}{S\arabic{page}}
\renewcommand{\thesection}{S\arabic{section}}
\renewcommand{\thetable}{S\arabic{table}}
\makeatletter

\renewcommand{\thesection}{\arabic{section}}
\renewcommand{\thesubsection}{\thesection.\arabic{subsection}}
\renewcommand{\thesubsubsection}{\thesubsection.\arabic{subsubsection}}

In this Supplemental Material, we discuss
\begin{enumerate}
    \item The numerical implementation based on free fermionic techniques;
    \item Finite-size scaling methods and additional numerical results.
\end{enumerate}

\section{Numerical methods}
This section briefly reviews the numerical methods employed in our work, cf.  Ref.~\cite{bravyi2004lagrangianrepresentationfor,turkeshi2022entanglementtransitionsfrom,alberton2021entanglementtransitionin,coppola2022growth,turkeshi2021measurementinducedentanglement} for more details. The efficiency of these techniques stems from the free fermionic circuits preserving the Gaussianity of an initial state. This reflects in an exponential simplification in the computational complexity, with the system's state fully encoded in the two-fermions correlation matrix.

\subsection{Majorana circuit}
The system is encoded in the Majorana correlation matrix $M_{a,b} = i\mathrm{tr}(\rho [\hat{\gamma}_a,\hat{\gamma}_b])/2$. The circuit dynamics stemming from $\hat{U}_{S\cup A}$, $\hat{\mathcal{U}}_S$ and $\hat{K}_p$ reflect in an effective evolution of $M$, cf. Ref.~\cite{bravyi2004lagrangianrepresentationfor}.

Every unitary operation, being quadratic in the Majorana fermions is representable as $\hat{U}_H(t) = \exp[i t\sum_{a,b}^{2N} H_{a,b} \hat{\gamma}_a\hat{\gamma}_b/4 ]$, with $N$ the total number of fermions in the system (plus ancillae), and $H$ some antisymmetric matrix with real entries. 
A simple exercise shows that any unitary transformation reflects in a rotation of the correlation matrix, namely
\begin{equation}
    M(t) = e^{-H t} M(0) e^{-H^T t}.\label{seq:hamdyn}
\end{equation}
In our Floquet implementation, $t=1$ and the energy scales of the dynamics are absorbed in $H$. 
Similarly, measurements of any Majorana parity -- the case of interest for our work, are efficiently computable. The following equation refers to the measurement over $i\hat{\gamma}_a\hat{\gamma}_b$, and stem from Wick's theorem
\begin{equation}
    M'_\pm  = \pm K + P\left(M+\frac{1}{1\pm M_{a,b}}MKM\right)P,\label{seq:mesdyn}
\end{equation}
with $M'_\pm$ the postmeasurement correlation matrix, $K_{p,q} = \delta_{p,a}\delta_{q,b}- \delta_{p,b}\delta_{q,a}$, $P = 1_N - \delta_{p,a}\delta_{q,a}+ \delta_{p,b}\delta_{q,b}$, $1_N$ the identity matrix over the $N$ degrees of freedom. The sign is chosen through the Born rule, with probabilities $p_\pm = (1\pm M_{a,b})/2$. 
Our numerical implementation follows Eqs.~\eqref{seq:hamdyn}-\eqref{seq:mesdyn}, replacing $H$ with the proper Ising-like gates (cf. Main Text) and with the projections on the local magnetization, i.e. $i\hat{\gamma}_{2i-1}\hat{\gamma}_{2i}$ with $i$ running through the system.

\subsection{Dirac circuit}
The system is encoded in the correlation matrix $C_{a,b} = \mathrm{tr}(\rho \hat{c}_a^\dagger \hat{c}_b)$. As for the Majorana case, the quadratic unitary gates and measurements reflect in a simplified dynamics of $C$, cf.~\cite{alberton2021entanglementtransitionin,coppola2022growth}. 
The unitary transformation preserving U(1) are given by $\hat{U}_H(t) = \exp(-i \sum_{a,b}^N H_{a,b} \hat{c}_a^\dagger \hat{c}_b)$, with $H$ an hermitian matrix. A simple computation leads to the correlation matrix evolution
\begin{equation}
    C(t) = e^{-i H t} C(0) e^{i H t}.\label{seq:hamdyn2}
\end{equation}
We consider measurements on the local density of particle $\hat{n}_k = \hat{c}^\dagger_k\hat{c}_k$. 
This translates via Wick's theorem to the correlation matrix after the measurement at site $k$
\begin{equation}
    [C']_{i,j} = \begin{cases}\displaystyle C_{i,j} + \delta_{i,k}\delta_{k,j} - \frac{C_{i,k} C_{k,j}}{C_{k,k}},& \text{ measurement on } \hat{n}_k\\
    \displaystyle C_{i,j} - \delta_{i,k}\delta_{k,j} - \frac{(\delta_{i,k}-C_{i,k})(\delta_{j,k}- C_{k,j})}{1-C_{k,k}},& \text{ measurement on } \hat\openone-\hat{n}_k.
    \end{cases}\label{seq:mesdyn2}
\end{equation}
Our numerical implementation follows Eqs.~\eqref{seq:hamdyn2}-\eqref{seq:mesdyn2} with the proper replacements of $H$.

\begin{figure}[t!]
    \centering
    \includegraphics[width=\textwidth]{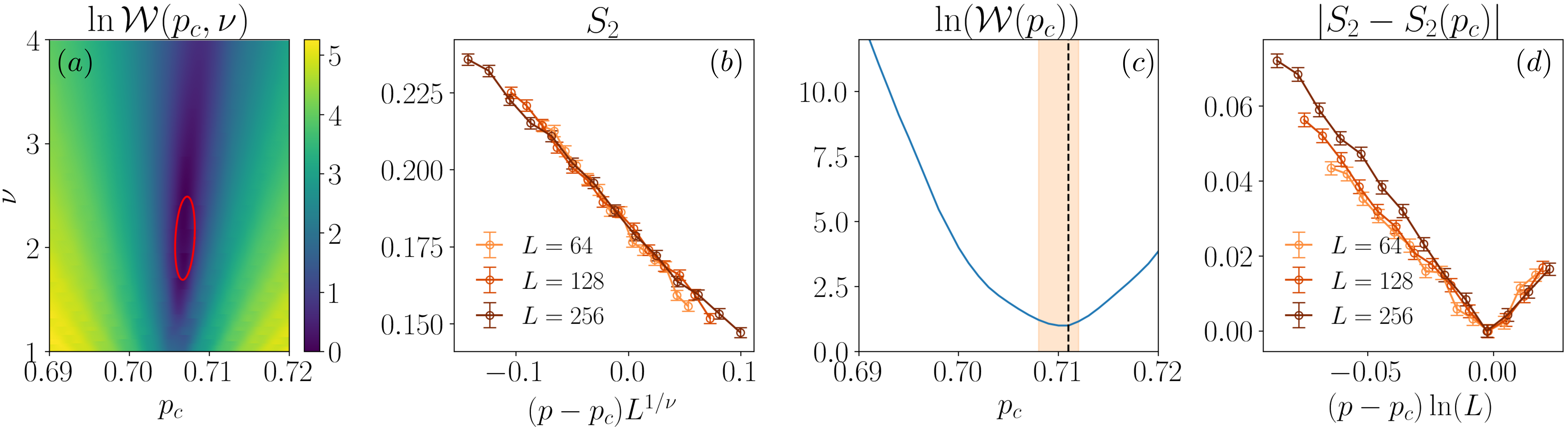}
    \caption{Majorana circuit finite size scaling. (a) The minimization landscape using a power-law scaling hypothesis locates the value reported in the Main Text $(p_c=0.707(3)$, $\nu=2.1(4))$. The red line denotes the interval of confidence, fixed by $\mathcal{W}<1.2$.
    To facilitate the comparison with the BKT scaling in (c,d), we recall the data collapse in Fig.3(c) (Main Text) for $S_2$. (c) The minimization landscape for the BKT scaling, locating $p_c=0.710(3)$. In Orange, the region for which $\mathcal{W}<1.2$. (d) Data collapse from the BKT finite size scaling. From the comparison of the collapses, we find the power-law diverging behavior better describes the data at hand. }
    \label{sfig:ising}
\end{figure}

\subsection{Choices of initial states}
We conclude by briefly summarizing the choices of initial states considered in our implementation. 

\subsubsection{Majorana circuit}
In the Main Text, we considered complementary but physically equivalent frameworks~\cite{block2022measurementinducedtransition}: the purification of an ancilla system initially entangled with the system and that of a maximally mixed initial state. 
Using the Main Text definitions, in the former case we choose $\hat{U}_{S\cup A}$ preparing the state described by the correlation matrix $M_{2i,2i+1} = - M_{2i+1,2i} = 1$ (with $2L+3\equiv 1$), and  $\hat{\mathcal{U}}_S=\hat{K}^L_{0,S}$ is $L$ applications of the purely unitary circuit ($p=0$). In conclusion, the initial state is
\begin{equation}
    \rho^\mathrm{majo}_{S\cup A}(0)=(\hat{K}^L_{0,S}\otimes\hat{\openone}_A) \left[\prod_{j=1}^{L+1} \frac{1+i\hat{\gamma}_{2j}\hat{\gamma}_{2j+1}}{2}\right](\hat{K}^L_{0,S}\otimes \hat{\openone}_A)^\dagger.
\end{equation}
Additionally, we have also considered $\hat{U}_{S\cup A}$ entangling the ancillae and the system locally starting from a product state $M_{2i-1,2i} =-M_{2i,2i-1}= 1$, and $\hat{\mathcal{U}}_S=\exp(i/4\sum_{a,b=1}^{2L}O_{a,b}\hat{\gamma}_a\hat{\gamma}_b)$ with $O$ a random antisymmetric matrix to test the robustness of our results. We find the qualitative purification timescales are unaltered, although we do not report these results for presentation purposes.  
In the second case, i.e. the study of the residual purification, the initial state is  $\rho_0=\openone/2^L$, with $M=0$ the zero matrix, and both $\hat{U}_{S\cup A}$ and $\hat{\mathcal{U}}_S$ being the identity operation.

\subsubsection{Dirac circuit}
Using the definitions in the Main Text and the fermionic occupations $n_i=0,1$, we consider the initial state obtained from $\ket{\Psi_{S\cup A}} = |0101\dots 01\rangle_S |0\rangle_A$ (described by $C_{i,j} = \delta_{2i,j}$), applying a local rotation $\hat{U}_{S\cup A}$ and $L$ scrambling purely unitary circuits $\hat{\mathcal{U}}_S=\hat{K}^L_{0,S}$. In summary, the initial state is
\begin{equation}
    |\Psi_{S\cup A}\rangle = (\hat{K}^L_{0,S}\otimes \hat{\openone}_A)\frac{|01\dots01\rangle_S|0\rangle_A+ |01\dots00\rangle_S|1\rangle_A}{\sqrt{2}}.
\end{equation}
Again, we have tested the numerical resilience varying initial conditions and found no qualitative difference in our results. 

\section{Finite-size scaling methods and additional numerical results}
This section discusses the finite-size scaling methods and additional numerical results concerning the robustness of our findings. 

We perform finite size scaling minimizing the cost function $W(\vec{a}) = \sum_{i=2}^{n-1} w(x_i,y_i,d_i)$ where $x_i$, $y_i$ and $d_i$ are respectively, the hyperparameters, data, and standard deviations that might depend on the optimization parameters $\vec{a}$. (Concretely $\vec{a}=(p_c,\nu)$ or $\vec{a} = (p_c)$, see below). 
The variables $x_i,y_i,d_i$ have index $i=0,\mathcal{N}_\mathrm{data}$ run over all the data considered and are ordered such as $x_i<x_j$ when $i<j$~\cite{Kawashima1993,zabalo2020criticalpropertiesof,sierant2022measurementinducedphase}. 
The density $w(x_i,y_i,d_i) = (y-\overline{y})^2/\Delta$ depends on $\overline{y} = \left[(x_{i+1}-x_i)y_{i-1} - (x_{i-1}-x_i)y_{i+1}\right] /(x_{i+1}-x_{i-1})$ and on 
\begin{equation}
    \Delta = d_i^2 + \left(\frac{x_{i+1}-x_{i}}{x_{i+1}-x_{i-1}}d_{i-1}\right)^2 + \left(\frac{x_{i-1}-x_{i}}{x_{i+1}-x_{i-1}}d_{i+1}\right)^2.\nonumber 
\end{equation}
Depending on the finite size scaling we consider, the value of $x_i,y_i,d_i$ is changed accordingly. The critical parameter are given minimizing $W(\vec{a})$, namely $\vec{a}^\star = \arg\min_{\vec{a}} W(\vec{a})$. We conveniently introduce the rescaled variable $\mathcal{W}(\vec{a}) = W(\vec{a})/W(\vec{a}^\star)\ge 1$, that we use to plot the landscape of critical parameters below. 
\begin{figure}[t!]
    \centering
    \includegraphics[width=0.8\columnwidth]{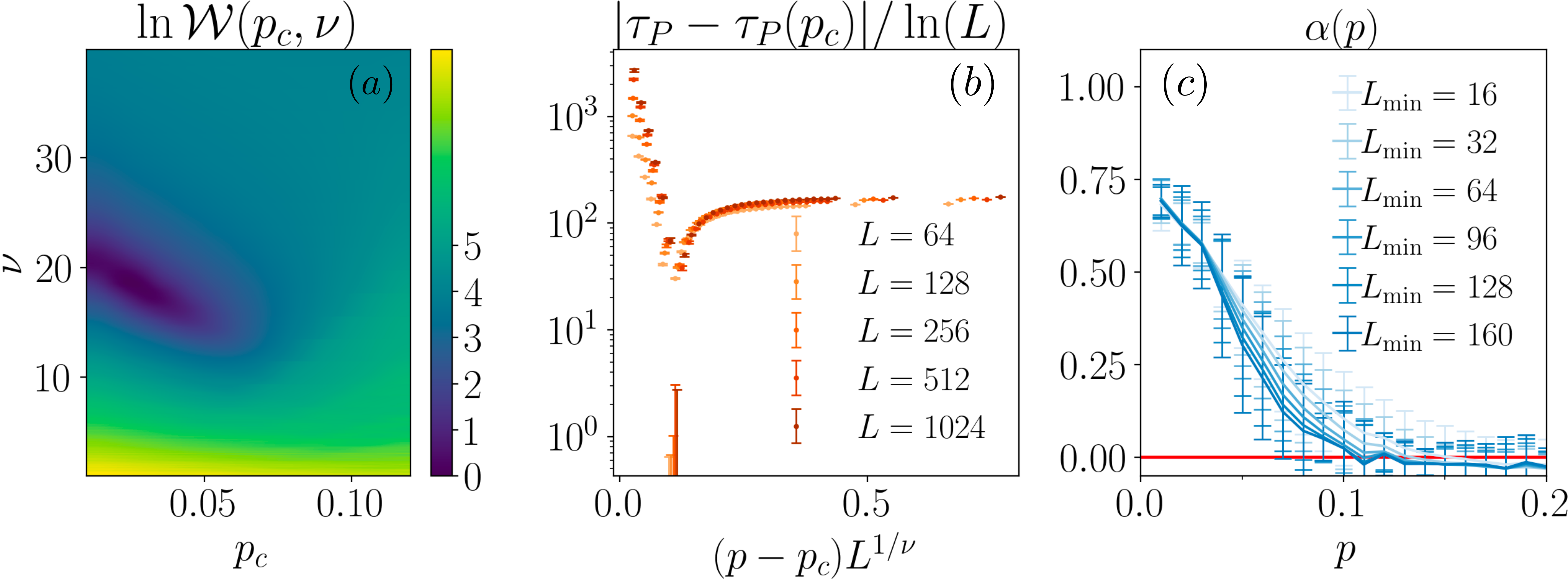}
    \caption{Dirac circuit finite size scaling considering a power-law scaling hypothesis. (a) The minimization landscape using a power-law scaling hypothesis locates the value reported in the Main Text $(p_c=0.020(5)$, $\nu=18(3))$. (b) Data collapse for $p_c=0.02,\nu=18$. The high value of $\nu$ is in line with the BKT scaling hypothesis.  }
    \label{sfig:u1}
\end{figure}
For both models, we consider the finite size scaling assuming a power-law divergence $x_i = (p_i-p_c)L^{1/\nu}_i$, $y_i=A_{i}$ and a BKT criticality $x_i = (p_i-p_c)\ln L$, $y_i=|A_{i}-A_i(p_c)|$, with $A$ the quantity of interest. 
For the Majorana circuit, we consider $A=S_2$ the residual entropy of an initially maximally mixed state $\rho=\openone/2^L$ after $t=L$ circuit layers. We specialize to $J=0.5$ and $h=0.3$, cf. Main Text.
The results are given in Fig.~\ref{sfig:ising}. The minimization landscape for a power-law divergence and BKT criticality are respectively given in Fig.~\ref{sfig:ising}(a), (c). We represent the interval of confidence as a red line, cf. Fig.~\ref{sfig:ising}(a), or an orange area, cf. Fig.~\ref{sfig:ising}(c). 
Comparing the data collapses  Fig.~\ref{sfig:ising}(b), (d), we find that the power-law divergence provides a better data collapse. This analysis concludes the transition is not of BKT type, as expected by the non-linear sigma model, cf. Main Text. 

For the Dirac circuit, we consider $A=\tau_P/\ln L$ the typical purification timescale. The BKT analysis was used in the main text, giving the collapse in Fig.3 (c) (Main Text). We complement this analysis by testing a power-law scaling hypothesis. 
The results are given in Fig.~\ref{sfig:u1}. From the minimization landscape, we find a high correlation length critical exponent $\nu\sim 20$, in line with the $\nu\to\infty$ expected from the BKT critical behavior. As we see, the quality of the data collapse, cf. Fig.~\ref{sfig:u1} is reasonable, albeit not as good as the BKT one presented in the text. 
Therefore, our findings rule out the power-law scaling behavior and support a BKT criticality, in line with previous results on continuously monitored systems~\cite{alberton2021entanglementtransitionin}. 

Lastly, in Fig.~\ref{sfig:u1}(c) we consider possible multiplicative logarithmic corrections in the mixed phase, i.e. $\tau_P\sim L^{\alpha(p)}\ln L$. As we see, compared to the Main Text results, the value of $\alpha(p)$ attains smaller values. The available numerics cannot differetiate this kind of fit from the purely power law presented in the Main Text, due to the double-logarithmic system size nature of the corrections.

\end{document}